\documentclass[reprint,prb,superscriptaddress,amsmath,amssymb,aps]{revtex4-1}

\usepackage{graphicx}
\usepackage{dcolumn}
\usepackage{bm}

\begin{document}


\title{Magnetotransport in phase-separated (Ga,Fe)N with $\gamma$'-Ga$_y$Fe$_{4-y}$N nanocrystals}


\author{A. Navarro-Quezada}
\email{andrea.navarro-quezada@jku.at}
\affiliation{Institute of Semiconductor and Solid-State Physics, Johannes Kepler University Linz, Altenberger Str. 69, 4040 Linz, Austria}

\author{M. Aiglinger}
\affiliation{Institute of Semiconductor and Solid-State Physics, Johannes Kepler University Linz, Altenberger Str. 69, 4040 Linz, Austria}

\author{B. Faina}
\affiliation{Institute of Semiconductor and Solid-State Physics, Johannes Kepler University Linz, Altenberger Str. 69, 4040 Linz, Austria}

\author{K. Gas}
\affiliation{Institute of Physics, Polish Academy of Sciences, Aleja Lotnikow 32/46, PL-02668 Warsaw, Poland}

\author{M. Matzer}
\affiliation{Institute of Semiconductor and Solid-State Physics, Johannes Kepler University Linz, Altenberger Str. 69, 4040 Linz, Austria}

\author{Tian Li}
\affiliation{Institute of Physics, Polish Academy of Sciences, Aleja Lotnikow 32/46, PL-02668 Warsaw, Poland}

\author{R. Adhikari}
\affiliation{Institute of Semiconductor and Solid-State Physics, Johannes Kepler University Linz, Altenberger Str. 69, 4040 Linz, Austria}

\author{M. Sawicki}
\affiliation{Institute of Physics, Polish Academy of Sciences, Aleja Lotnikow 32/46, PL-02668 Warsaw, Poland}

\author{A. Bonanni}
\email{alberta.bonanni@jku.at}
\affiliation{Institute of Semiconductor and Solid-State Physics, Johannes Kepler University Linz, Altenberger Str. 69, 4040 Linz, Austria}


\date{\today}

\begin{abstract}
The magnetotransport in phase-separated (Ga,Fe)N containing $\gamma$'-Ga$_y$Fe$_{4-y}$N (0\,$<$\,y\,$<$1) nanocrystals (NCs) is studied in the temperature range between 2\,K and 300\,K. The evolution of the resistivity and of the magnetoresistance (MR) as a function of temperature points at two conduction mechanisms: namely a conventional Arrhenius-type one down to 50\,K, and Mott variable range hopping at lower temperatures, where the spin-polarized current is transported between NCs in a regime in which phonon-scattering effects are not dominant. Below 25\,K, the MR shows a hysteretic contribution at magnetic fields $<$1\,T and proportional to the coercive field. Anisotropic magnetoresistance with values one order of magnitude greater than those previously reported for $\gamma$'-Fe$_4$N thin films over the whole considered temperature range, confirms that the observed MR in these layers is determined by the embedded nanocrystals.
\end{abstract}


\maketitle

\section{Introduction}

 The occurrence of nanospinodal decomposition is at the origin of the exceptional magnetic signatures in relevant semiconductors, like $e.g.$ Ge, GaAs, GaN and ZnO doped with magnetic elements and in particular with transition metal (TM) ions\,\cite{Bonanni:2010_CSR,Sato:2010_RMP,Dietl:2015_RMP}. In these material systems the aggregation of the TM ions
takes place either by preserving the crystallographic structure of the host lattice (chemical phase separation) or generating TM-rich nanocrystals (crystallographic phase separation) embedded in a TM-poor matrix. Crystallographic phase separation in particular, allows to combine the properties of the semiconducting host with those of embedded magnetic nanostrcutures, opening wide perspectives for spin detection and injection, spin valve effects, and flash-memory elements\,\cite{Coey:1999_JMMM}. 
 
 While the structural and magnetic properties of these phase-separated materials have been widely studied\,\cite{Dietl:2015_RMP}, unveiling comparable characteristics among the different systems, the understanding of the underlying (magneto)transport mechanisms is in its infancy. The transport characteristics and the magnetoresistance (MR) of semiconducting paramagnetic (PM) materials containing ferromagnetic nanocrystals (NCs)\,\cite{Akiyama:2012_APL} are dominated by spin and magnetic field dependent localization phenomena at the nanocrystal-matrix interface and significantly depend on the properties of the paramagnetic host, on the magnetic response of the embedded nanostructures and on the interaction of the electronic states of the host with the magnetic nanostructures.
 In (Ga,Mn)As phase-separated systems, a large negative MR associated with ferromagnetic MnAs nanoparticles (NPs) buried in the GaAs matrix was observed at temperatures below 30\,K\,\cite{Akinaga:1998_APL}, and a positive MR at intermediate fields up to 100\,K was reported\,\cite{Michel:2008_APL}. The large negative MR was explained in the frame of spin-dependent scattering of the carriers at the MnAs NPs, that decreases when the direction of the NPs magnetization aligns with the field. In (Ge,Mn) systems containing Ge$_3$Mn$_5$ ferromagnetic clusters and nanocolumns, it was found that the doping of the surrounding matrix plays a crucial role on the overall magnetotransport properties. In particular, a positive MR with a linear field dependence for layers grown on $p$-type GaAs was detected, while a negative MR with hysteretic behavior at 5\,K for layers grown on As-rich surfaces was reported\,\cite{Yu:2010_PRB}. The observed MR in this case was interpreted in terms of tunneling magnetoresistance effects. In ZnO containing Co embedded nanocrystals and at magnetic fields below 1\,T, a similar hysteretic behavior of the MR was also detected together with large positive MR values of 20\% at 10\,K\,\cite{Hamieh:2015_PRB}. In this material system, the transport properties are determined by the ZnO matrix containing magnetic localized Co$^{2+}$ impurities. Moreover, the hysteretic behavior indicates a small contribution to the conduction process at low fields, which is attributed to the spin polarization of the Co NCs. 

 In the case of GaN -- whose relevance for opto-electronics, high-power electronics and spintronics has steadily increased over the recent years -- crystallographic phase separation is observed when TM ions are incorporated above the solubility limit\,\cite{Giroud:2004_EPL,Kunert:2012_APL,Bonanni:2008_PRL,Bonanni:2007_PRB}. In particular, phase separation with ferromagnetic signatures up to 540\,K was reported in GaN doped with a concentration of Fe above 0.4\%\,\cite{Bonanni:2007_PRB} cations and grown by metalorganic vapour phase epitaxy (MOVPE). While the inhomogeneous distribution of a variety of ferromagnetic (FM) and antiferromagnetic (AF) Fe$_x$N phases $(x = 2,3,4)$ in this particular material system has to-date limited its applicability into functional devices\,\cite{Navarro:2010_PRB,Navarro:2011_PRB}, the recently reported spatial localization of arrays of single-phase face-centered cubic $\gamma$'-Ga$_y$Fe$_{4-y}$N NCs with in-plane uniaxial magnetic anisotropy has opened new perspectives for these material system\,\cite{Navarro:2012_APL,Grois:2014_Nanotech}. 
  
  Furthermore, by varying the fabrication conditions, the lattice parameter of the embedded $\gamma$'-Ga$_y$Fe$_{4-y}$N NCs can be tuned on-demand from the one of $\gamma$'-Fe$_4$N, towards the one of $\gamma$'-GaFe$_3$N by controlling the incorporation of Ga into the $\gamma$'-Fe$_4$N lattice. Since the magnetic response of $\gamma$'-Ga$_y$Fe$_{4-y}$N ranges from strongly ferromagnetic ($y$$<$0.25) to weakly antiferromagnetic ($y$$>$0.25)\,\cite{Houben:2009_ChemMat, Burghaus:2011_JSSC}, planar arrays of $\gamma$'-Ga$_y$Fe$_{4-y}$N NC arrays embedded in GaN become suitable for FM as well as for the emerging field of AF-spintronics\,\cite{Jungwirth:2016_NatNano, Wadley:2016_Science}. 

 Here, the magnetotransport mechanisms in MOVPE phase-separated (Ga,Fe)N thin layers (Ga$\delta$FeN) grown on GaN buffers and containing $\gamma$'-Ga$_y$Fe$_{4-y}$N nanocrystals are investigated in the temperature range between 2\,K and 300\,K. While the main conduction channel is confirmed to be the GaN buffer layer, the magnetoresistance of the system is significantly affected by the presence of the NCs, as evidenced by anisotropic magnetoresistance (AMR) observed at all temperatures. In consideration of the relevance of capping layers and encapsulation for applications, the behavior of GaN-capped phase-separated Ga$\delta$FeN is also analysed.

\section{Experimental details}

\begin{figure}
	\includegraphics[width=\columnwidth]{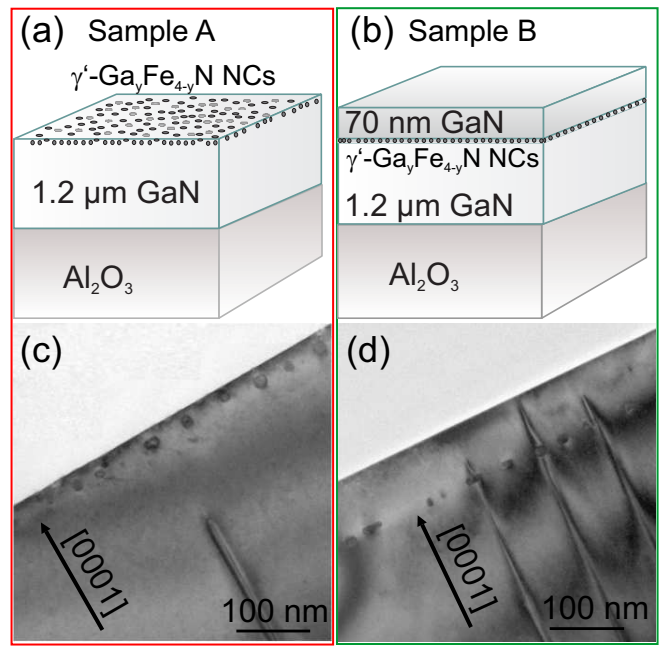}
	\caption{(a) Schematic representation of Sample A -- uncapped; (b) schematic representation of Sample B -- capped with a nominally 70\,nm thick GaN layer. (c) and (d): Cross-section TEM images of Samples A and B, respectively, evidencing the spatial distribution of the NCs in a plane perpendicular to the growth direction [0001].}
	\label{fig:struct1}
\end{figure}

The investigated samples are fabricated by MOVPE according to the procedure previously reported\,\cite{Navarro:2012_APL} and consist of a 1200\,nm thick GaN buffer layer -- unintentionally $n$-type with a carrier concentration $n$\,=\,(8$\times$10$^{16}$)\,cm$^{-3}$ at room-temperature (RT) -- deposited at 1040$^{\circ}$C on 2"\,$c$-[0001] sapphire substrates, followed by the growth of a 50\,nm thin Ga$\delta$FeN layer containing nanocrystals deposited at 780$^{\circ}$C, eventually capped with GaN. Specifically, the following structures are considered: (i) Sample A -- uncapped; (ii) Sample A* -- $i.e.$ Sample A upon annealing at 600$^\circ$C under N$_2$ carried out to remove the $\alpha$-Fe inclusions formed in proximity of the sample surface when the samples are left uncapped; (iii) Sample B -- with the same basic structure as Samples A and A*, but additionally capped with a nominally 70\,nm thick GaN layer grown at 1000$^{\circ}$C. A schematic representation of the samples architecture is reproduced in Figs.\,\ref{fig:struct1}(a) and (b). A bare 1200\,nm thick GaN buffer layer on sapphire is employed as reference. The relevant characteristics of the samples are provided in Table\,I.

 Information on the samples structure and on the NCs crystallographic phase is obtained from high-resolution x-ray diffraction (HRXRD) rocking curves. The measurements are carried out using a PANalytical X’Pert Pro Material Research Diffractometer equipped with a hybrid monochromator with a 0.25$^{\circ}$ divergence slit and a PixCel detector using 19 channels for detection and a 5\,mm anti-scatter slit. In order to determine the distribution of the NCs in the GaN matrix, as well as to verify the crystallographic phase of the NCs, (high-resolution) transmission electron microscopy (HRTEM) measurements are performed. The cross-section TEM specimens are prepared by mechanical polishing, dimpling and ion milling in a Gatan Precision Ion Polishing System. Dark-field and bright-field measurements in conventional mode, and high-angular annular dark-field (HAADF) measurements in STEM mode are performed using a JEOL JEM\,2000\,EX system. 

The magnetic characteristics of the samples are measured in a Quantum Design superconducting quantum interference device (SQUID) magnetometer MPMS XL. The samples are investigated in the temperature range between 2\,K and 300\,K in magnetic fields up to 5\,T. The diamagnetic component originating from the sapphire substrate is carefully compensated according to the procedure described elsewhere\,\cite{Sawicki:2011_SST,Gas:2018_arxiv}.
Additionally, the angular dependence of the magnetization is studied by performing ferromagnetic resonance (FMR) measurements at RT in out-of-plane and in-plane configuration with a Bruker Elexsys E580 electron paramagnetic resonance spectrometer at microwave frequencies between 9.4\,GHz and 9.5\,GHz. During measurements, the static magnetic field is modulated with an amplitude of 0.5\,mT at 100\,kHz to allow lock-in detection\,\cite{Grois:2014_Nanotech}. 
 
The magnetotransport measurements are carried out in Van der Pauw geometry using a high-sensitivity Janis Super Variable Temperature 7TM-SVM magnetotransport system in the temperature range between 2\,K and 300\,K and in an external magnetic field $H$ tunable between -6\,T and 6\,T, applied perpendicular to the film plane.  Details on the fabrication of the contacts are provided in the Supplemental Material\,\cite{supplementary}. For AMR measurements, the direction of $H$ has been varied with respect to the surface sample normal from perpendicular (out-of-plane angle $\beta = 90^\circ$) to parallel ($\beta = 0^\circ$).

\section{Results and Discussion}
\subsection{Structural and magnetic properties}{\label{3a}}

\begin{figure}
	\includegraphics[width=\columnwidth]{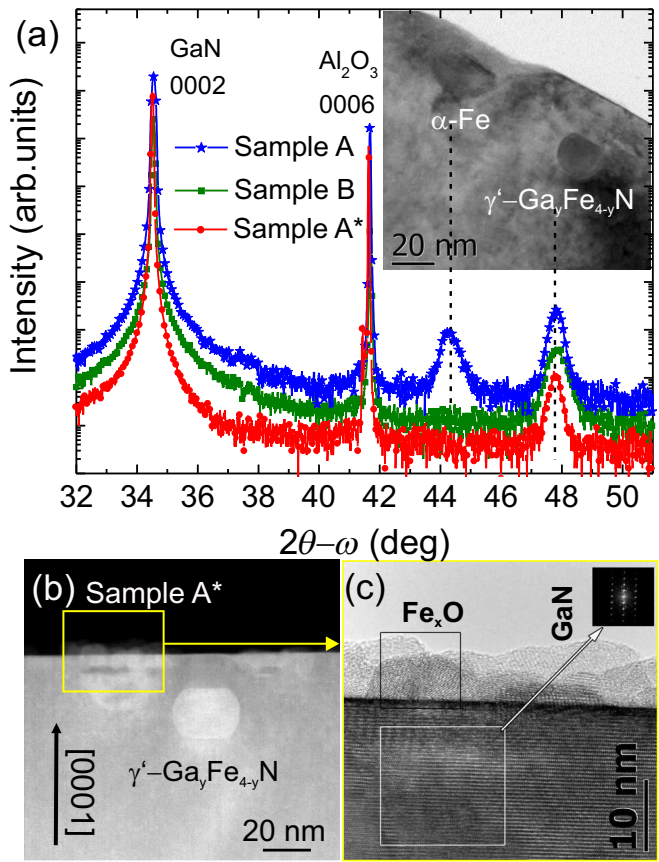}
	\caption{(a) HRXRD of the samples with the identified NCs diffraction peaks. Inset to (a): HRTEM of $\alpha$-Fe and $\gamma$'-Ga$_y$Fe$_{4-y}$N in Sample A. (b) HAADF cross-section TEM image of Sample A*. (c) HRTEM of the polycrystalline FeO$_x$ formed at the surface of the area occupied by an $\alpha$-Fe NC prior to annealing of Sample A. Inset to (c): SAD pattern of GaN in proximity of the area previously occupied by $\alpha$-Fe, confirming the outward diffusion of Fe.}
	\label{fig:xrd1}
\end{figure}

 The structure of the nanocrystals present in the Ga$\delta$FeN layers is assessed by HRXRD: besides the 0002 diffraction peak of the GaN buffer and the 0006 peak of the Al$_2$O$_3$ substrate, two additional reflections are observed in the spectra reported in Fig.\,\ref{fig:xrd1}(a) for Sample A at 44.275$^{\circ}$$\pm$0.005$^{\circ}$ and 47.821$^{\circ}$$\pm$0.005$^{\circ}$. In Sample B the peak at around 47.8$^{\circ}$ solely is detected. From the distance between the diffraction planes ($d$-spacing), the reflections are identified as the (110) of $\alpha$-Fe and the (200) of $\gamma$'-Ga$_y$Fe$_{4-y}$N, respectively\,\cite{Rovezzi:2012_arxiv}. These two crystalline phases were already reported for phase-separated (Ga,Fe)N\,\cite{Navarro:2010_PRB, Li:2008_JCG}. From the full-width-at-half-maximum (FWHM) of the diffraction peaks and by employing the Scherrer formula\,\cite{Patterson:1939_PR}, the size of the nanocrystals along the growth direction is estimated to be (17\,$\pm$2)\,nm for the $\alpha$-Fe and (22\,$\pm$2)\,nm for the $\gamma$'-Ga$_y$Fe$_{4-y}$N nanocrystals.

The presence of the two crystallographic phases in Sample A is confirmed by the cross-section HRTEM image reported in the inset to Fig.\,\ref{fig:xrd1}(a): $\gamma$'-Ga$_y$Fe$_{4-y}$N nanocrystals buried 30\,nm below the sample surface and -- as identified by selective area diffraction pattern (SAD) analysis -- $\alpha$-Fe NCs located at the surface. It is estimated that this phase accounts for $\sim$50\% of the total NCs in the sample. The tendency of the $\alpha$-Fe to form at the sample surface was previously reported and attributed to the evaporation of nitrogen taking place during the MOVPE process\,\cite{Li:2008_JCG}. According to the cross-section image in Fig.\,\ref{fig:struct1}(d), in Sample B only embedded $\gamma$'-Ga$_y$Fe$_{4-y}$N are found located 70\,nm to 100\,nm below the sample surface, in agreement with the nominal thickness of the GaN capping layer. The NCs size obtained from TEM images ranges between 10\,nm and 30\,nm, consistent with the average size obtained from the FWHM analysis of the HRXRD spectra. Energy dispersive x-ray (EDX) scattering performed during the TEM imaging, points at a minimal concentration $x$$<$0.1\% of dilute Fe in the host matrix, suggesting that most of the Fe provided during growth is incorporated into the nanocrystals. The dilute Fe ions in the GaN matrix are known to occupy preferentially substitutional Ga-sites and be in the Fe$^{3+}$ charge state\,\cite{Bonanni:2007_PRB}. 

The Fe-rich NCs embedded in previously considered phase-separated (Ga,Fe)N layers were found to be stable up to temperatures as high as 900$^\circ$C\,\cite{Navarro:2010_PRB}. This behavior is confirmed for Sample B, in which the NCs are buried below the GaN capping layer, but in the case of Sample A, post-growth annealing carried out at 600$^\circ$C under N$_2$ atmosphere -- and resulting in Sample A* -- significantly affects the system, as detailed in Fig.\,S1(a) of the Supplemental Material\,\cite{supplementary}. The XRD analysis of Sample A* ($i.e.$ Sample A annealed) points at a removal of $\alpha$-Fe NCs upon annealing, as evidenced in Fig.\,\ref{fig:xrd1}(a). This effect is corroborated by the detailed TEM analysis of Sample A* summarized in Figs.\,\ref{fig:xrd1}(b) and (c). In the HAADF cross-section image of Fig.\,\ref{fig:xrd1}(b) a $\gamma$'-Ga$_y$Fe$_{4-y}$N NC as well as a void resulting from the removal of $\alpha$-Fe and located at the sample surface are pictured. In the HRTEM image in Fig.\,\ref{fig:xrd1}(c), the region adjacent to the void is considered in detail and a SAD analysis points to the structure of GaN. Moreover -- again upon annealing -- the presence of polycrystalline FeO$_x$ at the surface is detected, suggesting out-diffusion of the Fe originally bound in $\alpha$-Fe NCs. The polycrystalline FeO$_x$ is completely removed through chemical treatment of 1 hour in HCl, as confirmed by x-ray photoemission spectroscopy (XPS) and as evidenced in Fig.\,S1(b) of the Supplemental Material \,\cite{supplementary}. 

The magnetic response of the Ga$\delta$FeN layers investigated here is qualitatively similar to the one of phase-separated (Ga,Fe)N layers previously reported\,\cite{Bonanni:2007_PRB}.
The field dependency of the in-plane aerial density of magnetic moment $m/m_\mathrm{s}$ is characterized primarily by a strongly non-linear superparamagnetic-like response detected over the whole range of the studied temperatures and by a PM contribution from dilute substitutional Fe$^{3+}$ ions in the matrix dominating at low temperatures. These two magnetic components can be treated separately and quantitatively, as previously reported\,\cite{Pacuski:2008_PRL,Navarro:2010_PRB}.

\begin{figure}
	\includegraphics[width=\columnwidth]{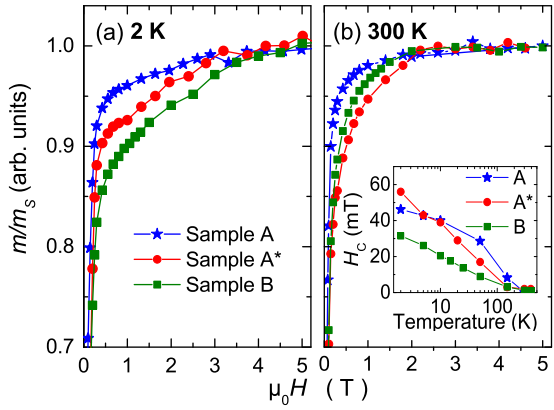}
	\caption{Normalized aerial density of magnetic moment of Samples A (stars), A* (dots) and B (squares): (a) at 2\,K and (b) at 300\,K. Inset to (b): Coercive field strength ($H_c$) \textit{vs.} temperature.}
	\label{fig:magnetic}
\end{figure}

 In Figs.\,\ref{fig:magnetic}(a) and (b), the normalized aerial density of magnetic moment as a function of the applied magnetic field is compared for the considered samples containing NCs, both at 2\,K and at 300\,K, respectively.  In Sample A the magnetization saturates swiftly at 1\,T at room-temperature, as expected due to the presence of the soft ferromagnetic $\alpha$-Fe, known to saturate at fields as low as 10\,mT at RT\,\cite{Nie:2003_TSF}. Upon annealing -- $i.e.$ Sample A* -- the response quenches and its saturation shifts from 2\,T to 5\,T at 2\,K, as evidenced in Fig.\,\ref{fig:magnetic}(a) and from 1\,T to 2\,T at RT, as reported in Fig.\,\ref{fig:magnetic}(b). 
 
 All samples show an open hysteresis below 100\,K with coercive field values between 10\,mT and 60\,mT. The coercive field for $\alpha$-Fe is reported to be as low as 0.2\,mT at RT\,\cite{Nie:2003_TSF}, while the one of $\gamma$'-Fe$_4$N nanoparticles is in the range between 50\,mT and 80\,mT\,\cite{Wu:2004_JMMM}. These last values fit well to the ones observed in the considered samples and evidenced in the inset to Fig.\,\ref{fig:magnetic}(b), pointing at the presence of FM $\gamma$'-Fe$_4$N nanocrystals. 
 
 Taking into consideration the structural information previously discussed, the difference in magnetization between Samples A and A* is calculated. The difference signal is nearly temperature independent -- as detailed in Fig.\,S2 of the Supplemental Material\,\cite{supplementary} -- rapidly saturates with increasing $H$ and its saturation as a function of temperature is comparable to the one observed for Fe nanoparticles\,\cite{Zhang:1998_PRB}, hinting at ferromagnetic $\alpha$-Fe NCs as main responsible for the difference between the two responses. Remarkably, the amount of Fe$^{3+}$ ions contributing to the overall magnetic signal is the same for both samples, indicating that dilute Fe in the host matrix is not affected by the annealing procedure, in accordance with previous findings\,\cite{Navarro:2010_PRB}.

On the other hand, the normalized magnetization as a function of the applied field is comparable for Samples B and A* at 2\,K, while significantly differing at RT: both specimens are found to contain only $\gamma$'-Ga$_y$Fe$_{4-y}$N (and not $\alpha$-Fe) NCs, but with density, as well as composition ($y$), differing in the two samples. 

\begin{figure}
	\includegraphics[width=\columnwidth]{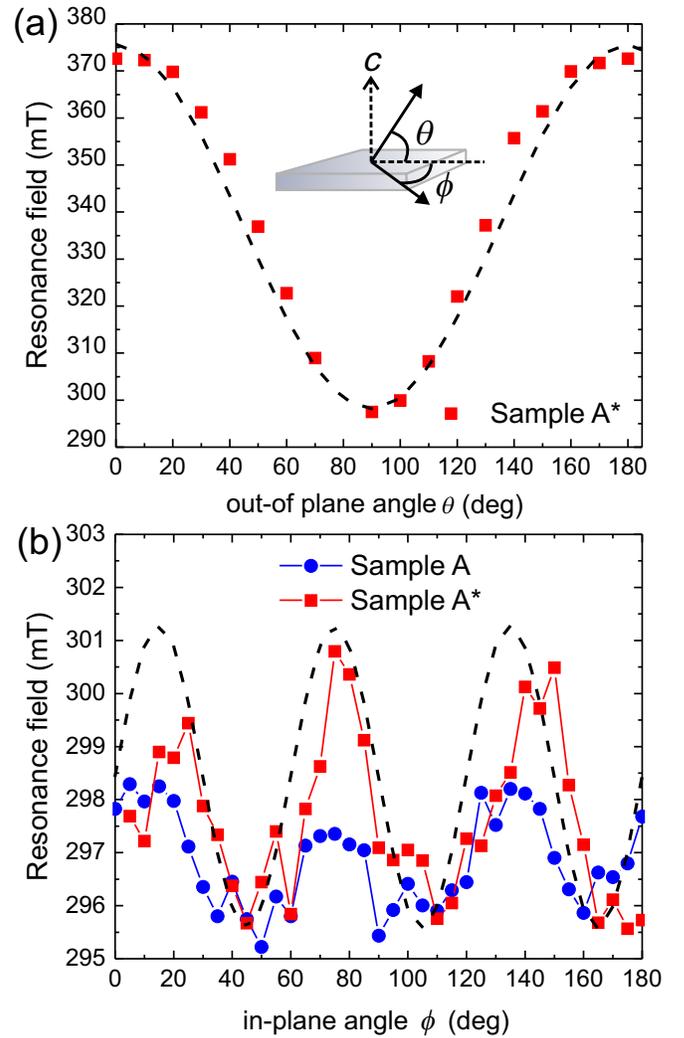}
	\caption{ Angular dependence of the FMR signal for Samples A and A*: (a) the out-of-plane FMR follows a cos$^2\theta$ dependence, while (b) the in-plane signal follows a sin(6$\phi$) behavior.}
	\label{fig:FMR}
\end{figure}
 
The out-of-plane and in-plane angular dependence of the FMR signals provided in Fig.\,\ref{fig:FMR} confirm that the magnetic anisotropy of Sample A is equivalent to the one observed for capped samples\,\cite{Grois:2014_Nanotech}. An uniaxial $cos^2\theta$ dependence of the resonance field is observed when varying the out-of-plane angle ($\theta$), as evidenced in Fig.\,\ref{fig:FMR}(a), while the variation of the in-plane angle (azimuth $\phi$) produces a $sin(6\phi)$ dependence reported in Fig.\,\ref{fig:FMR}(b). The uniaxial out-of plane anisotropy is attributed to shape anisotropy, while the in-plane anisotropy is related to the hexagonal crystal structure of the GaN matrix. The FMR results point at three easy axes lying in the plane of the NCs. 

\begin{table*}[ht]
	\centering
	\begin{tabular}{c| c c c c c c c}
		\hline \hline
		Sample & Structure & Annealing & $R_\mathrm{RT}$  		& $n$ 									&  $\mu$  			& $E_\mathrm{a}$ 	& $T_{0}^{\mathrm{Mott}}$ \\
					 &           &           & $(\mathrm{k}\Omega$) 	& $(10^{16}$ cm$^{-3}$) &  (cm$^2$/Vs) 	& (meV) 		& (K) \\
		\hline
		GaN & 1200$\mu$m GaN layer & none & 3 & 8 & 350 & 20 & $7\times10^{4}$ \\
		A & uncapped Ga$\delta$FeN/GaN & none & 5 & 6 & 600 & 17 & $1\times10^{7}$ \\
		A* & uncapped Ga$\delta$FeN/GaN & 600$^\circ$C/N$_{2}$/10min & 6 & 8 & 700 & 19 & $6\times10^{6}$ \\
		B & capped Ga$\delta$FeN/GaN & none & 43 & 2 & 100 & 25 & $6\times10^{12}$ \\
		\hline \hline
	\end{tabular}
	\label{tab:tab1}
	\caption{Structure and characteristic transport parameters of the investigated samples. The mobility $\mu$, the carrier density $n$ and the resistance $R$ values are acquired at RT. With $R$ the average resistance across the four ohmic contacts in van der Pauw geometry). The activation energy $E_\mathrm{a}$ and the $T_{0}^{\mathrm{Mott}}$ are obtained from fitting $\rho(T)$.}
\end{table*}

\subsection{Resistivity and magnetoresistance}

\begin{figure}
	\includegraphics[width=0.9\columnwidth]{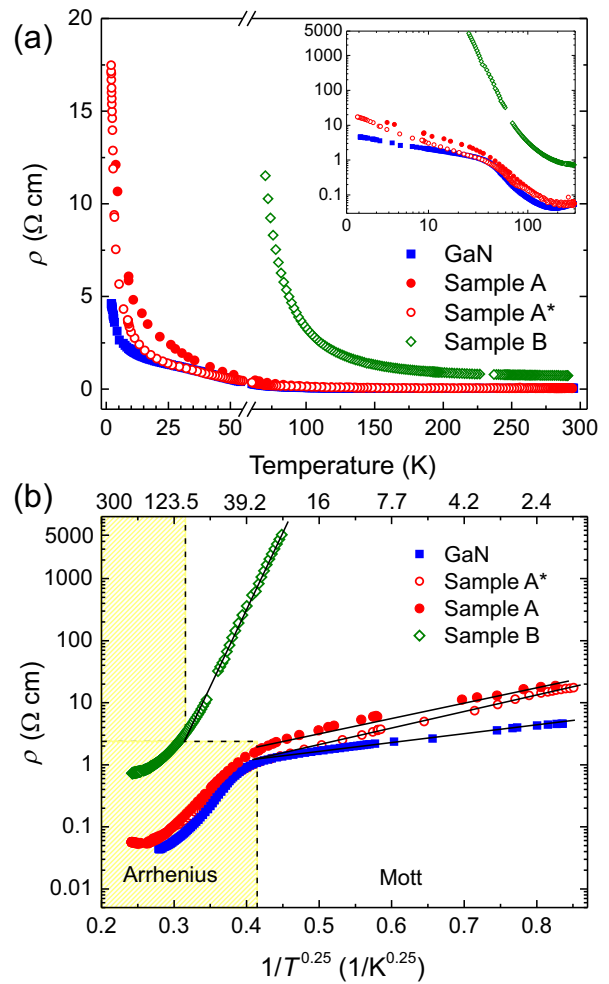}
	\caption{(a) Resistivity \textit{vs.} temperature for Samples A, A* and B compared to GaN. (b) Resistivity \textit{vs.} $T^{0.25}$ pointing at a characteristic Mott VRH type conduction below 40\,K for Sample A, for Sample A* and for the GaN reference, and for Sample B below 130\,K. Above these temperatures, an Arrhenius-like behavior is observed.}
	\label{fig:rho}
\end{figure}

For completeness and considering that the Ga$\delta$FeN are grown on GaN buffer layers with unintentional $n$-conductivity, the bare reference GaN buffer is also studied. The resistivity $\rho$ of the samples is measured as a function of temperature and reported in Fig.\,\ref{fig:rho}(a). The characteristic transport parameters obtained at RT for all samples are listed in Table I. 

The resistivity of all samples increases with decreasing temperature, as expected for semiconducting thin films. While the $\rho$s of Sample A, of Sample A* and of the GaN reference are comparable down to 50\,K, the one of Sample B is enhanced by approximately one order of magnitude at temperatures down to 100\,K and is about three orders of magnitude larger below this temperature. In the inset to Fig. 5(a) the $\rho(T)$ is depicted in full scale, evidencing the large difference in resistivity between Sample B and the other measured samples. This effect is attributed to the diffusion of dilute Fe ions -- with a concentration $x$$\sim0.1$\% -- from the Ga$\delta$FeN region into the GaN capping layer of Sample B, which compensates the unintentional $n$-type conductivity of the GaN capping layer\,\cite{Heikman:2002_APL} and increases the measured resistivity.  

The temperature dependence of $\rho (T)$ points at two distinct regimes. Specifically, an Arrhenius-type conduction is observed for Sample A, for Sample A* and for the GaN reference down to 50\,K, and for Sample B down to 130\,K. The activation energies $E_\mathrm{a}$ for all samples are comparable, suggesting that the transport in this conduction regime takes place mainly in the underlying GaN buffer. The obtained $E_\mathrm{a}$ values fit well with those previously reported for GaN layers\,\cite{Goetz:1996_APL,Yildiz:2010_APA}. Furthermore, the behavior of $\rho(T)$ points at a Mott variable range hopping (VRH) conduction mechanism, as evidenced in Fig.\,\ref{fig:rho}(b), where the linear dependency of $\rho(T)$ on $T^{-0.25}$ is reported. The transition from Arrhenius to Mott VRH conduction occurs between 50\,K and 25\,K in Sample A, Sample A* and GaN, while it shifts to between 130\,K and 100\,K for Sample B.

The VRH electronic transport is characterized by conduction in an impurity band of localized states with random spatial and energy distribution\,\cite{Pollak:2002_pssb} and a constant density of states at the Fermi energy $N(E_\mathrm{F})$. In the Mott VRH model the resistivity as a function of temperature follows\,\cite{Mott:1969}:

\begin{equation}
\rho (T) \sim exp(T_{0}^{\mathrm{Mott}}/T)^{1/4},
\end{equation}
\label{eq:one}

where $T_{0}^{\mathrm{Mott}}$ is the characteristic hopping temperature, whose value depends on the electronic density of states at the Fermi level and on the localization length $\xi$. 
 The linear fits in the Mott regime yield the $T_{0}^{\mathrm{Mott}}$ values listed in Table I, that are in good agreement with those previously reported for Gd-doped GaN\,\cite{Pinto:2009_PRB} and for GaN\,\cite{Yildiz:2010_APA}. The value obtained for $T_{0}^{\mathrm{Mott}}$ in the GaN reference is two orders of magnitude lower than those for Samples A and A*, suggesting a larger ratio between the hopping distance $R_\mathrm{h}$ and the localization length $\xi$ in the uncapped Ga$\delta$FeN layers, likely to be related to the presence of $\gamma$'-Ga$_y$Fe$_{4-y}$N NCs and of dilute Fe ions in these layers. Sample A and Sample A* show a comparable behavior and correspondent fitting parameters, indicating that the $\alpha$-Fe NCs present in Sample A give a negligible contribution to the transport. One way to determine the values of the localization length and of the hopping energy from the obtained $T_{0}^{\mathrm{Mott}}$, goes through the knowledge of the density of states at the Fermi level. Alternatively, these values can be obtained from magnetoresistance measurements, as shown below.

\begin{table}[ht]
	\centering
	\begin{tabular}{c| c c c c}
		\hline \hline
		& GaN &         &    Sample A* & \\
		\hline
		$T$(K) & $a\pm 0.05$ & $b\pm 0.05$ & $a\pm 0.05$ & $b\pm 0.05$ \\
		\hline
		50 & $0.42$ & 0.00 & $0.36$ & $0.04$ \\
		100 & $0.63$ & 0.00 & $0.58$ & $0.06$ \\
		150 & $0.65$ & 0.00 & $0.61$ & $0.07$ \\
		200 & -  & -  & $0.47$ & $0.07$ \\
		250 & $0.37$ & 0.00 & $0.38$ & $0.05$ \\
		\hline \hline
	\end{tabular}
	\label{table2}
	\caption{Two-band model\,\cite{Zaremba:1992_PRB} fit parameters for the MR above 50\,K in the GaN reference and in Sample A*.}
\end{table}

\begin{figure}
	\includegraphics[width=\columnwidth]{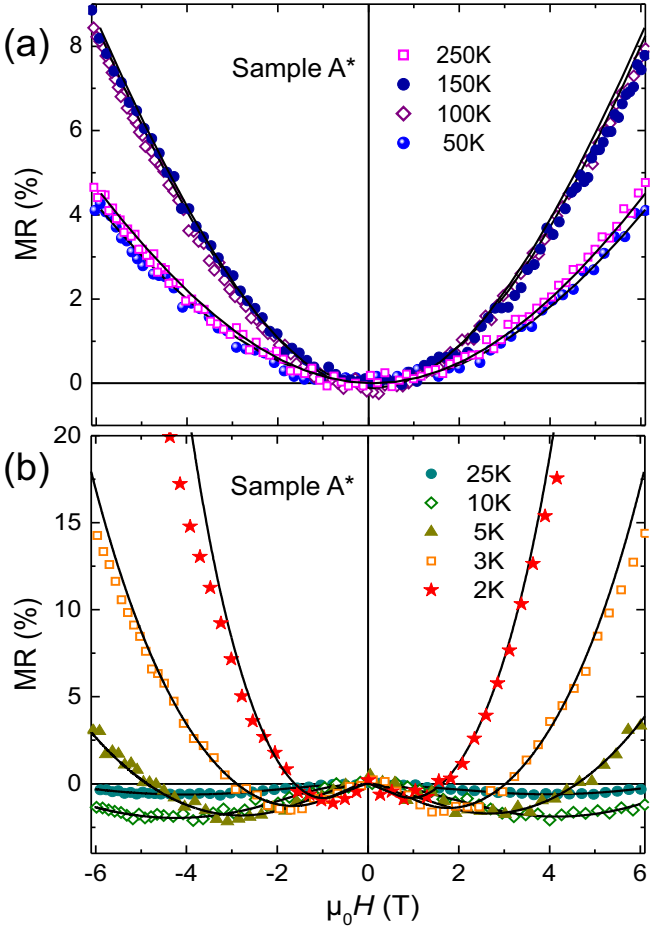}
	\caption{Magnetoresistance of Sample A* as a function of the applied magnetic field: (a) above 50\,K, and (b) below 50\,K. Symbols: experimental data; solid lines: fits.}
	\label{fig:MR}
\end{figure}

In order to gain insight into the contribution of the NCs to the (magneto)transport in the studied structures, the resistance of the layers is measured as a function of the applied magnetic field. The MR is defined as:

\begin{equation}
\frac{\Delta\rho}{\rho_0} = \frac{\rho(H,T)-\rho_{0}(T)}{\rho_{0}(T)},
\end{equation}

where $\rho_0$ and $\rho(H)$ are the resistivity at $H$=0 and at a field $H$$\neq$0, respectively. In Figs.\,\ref{fig:MR}(a) and (b) the MR of Sample A* as a function of the applied magnetic field is reported: as evidenced in Fig.\,\ref{fig:MR}(a), at temperatures $\geq$50\,K the MR is positive and follows a $H^2$ dependence similar to the one of the GaN reference\, presented in Fig.\,S3 of the Supplemental Material\,\cite{supplementary}. The positive MR can be explained in the frame of a two-band conduction model, in which one conduction band is the one of GaN and the second one is an impurity band induced by the diluted Fe ions in the host matrix. In this model, the MR is given by:

\begin{equation}
	\mathrm{MR}=a^2(\mu_0H)^2/[1+b^2(\mu_0H)^2],
\end{equation}

 where $\mu_0$ is the vacuum permeability, while $a$ and $b$ are parameters directly related to the conductivities and mobilities of each band\,\cite{Zaremba:1992_PRB}. The values of $a$ and $b$ obtained by fitting with the mentioned model the MR of Sample A* for temperatures $\geq$50\,K are collected in Table II. While for GaN the MR can be fitted considering one single band, two bands are required for Sample A*. The values of the fit parameter $a$ are comparable to those found for the GaN reference, confirming that the dominating conduction channel at high temperatures in this sample is the underlying GaN buffer.

The MR curves acquired at temperatures below 50\,K are represented in Fig.\,\ref{fig:MR}(b), and show a negative slope up to a critical field $H_\mathrm{s}$, and a positive quadratic behavior at higher magnetic fields. In the VRH regime, the MR depends on the variations in the hopping probability with the magnetic field. Specifically, for low magnetic fields, the negative magnetoresistance (NMR) is due to quantum interference of different hopping paths between initial and final states, similarly to weak-localization in metals\,\cite{Zhao:1991_PRB}. When the magnetic field is increased, the effect is overcome by the presence of a strong positive MR\,\cite{Gantmakher:1996_JETP}. The critical value of the field at which the MR changes from negative to positive allows obtaining directly the characteristic hopping parameters $R_\mathrm{h}$ and $\xi$\,\cite{Entin:1989_PRB}. 

\begin{figure}
	\includegraphics[width=\columnwidth]{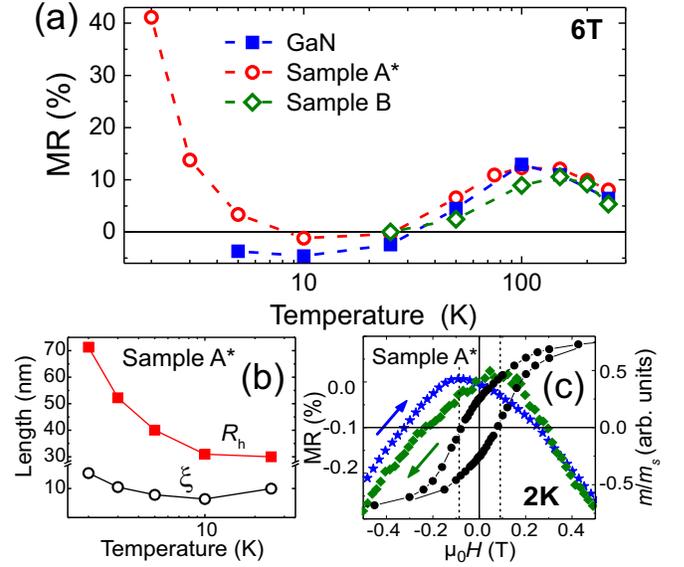}
	\caption{(a) Comparison of the MR as a function of temperature for Samples A, A*, GaN and B at 6\,T. (b) $R_\mathrm{h}$ and $\xi$ obtained as a function of temperature for Sample A*. (c) MR and normalized density of magnetic moment as a function of the applied field for Sample A*. The field sweep directions are indicated by arrows.}
	\label{fig:MR2_1}
\end{figure}

The evolution of the MR acquired at 6\,T as a function of temperature for all samples is reported in Fig.\,\ref{fig:MR2_1}(a). At temperatures above 100\,K the MR of all samples is positive and follows a $H^2$ behavior, while at lower temperatures the MR of GaN is negative, and the one of Sample A* is dominantly positive reaching 42\% at 2\,K. The NMR of the GaN reference layer is analyzed in the frame of the Mott VRH conduction\,\cite{supplementary}, similarly to the case of \textit{n}-type InP\,\cite{Abdia:2009_SSE}. 
The low temperature MR of Sample A* is fitted according to the model developed by Nguyen, Spivak and Shlovskii\,\cite{Zhang:1992_PRB}, and $R_\mathrm{h}$ and $\xi$ as a function of temperature directly obtained from the critical field $H_\mathrm{s}$ are plotted in Fig.\,\ref{fig:MR2_1}(b). The values of $R_\mathrm{h}$ lie between 30\,nm and 70\,nm, while those of $\xi$ vary in the range (7-10)\,nm, fulfilling the condition $R_\mathrm{h}$$>$$\xi$ for hopping conduction\,\cite{Mott:1969}. From HRTEM plane-view measurements\,\cite{Navarro:2012_APL}, the average distance between the NCs is found to range between 30\,nm and 100\,nm, rendering hopping conduction through the NCs significant at temperatures below 10\,K.  

As evidenced in Fig.\,\ref{fig:MR2_1}(c), a hysteretic behavior of the MR of Sample A* is observed at low fields, similar to the one reported for the ZnO:Co phase-separated system\,\cite{Hamieh:2015_PRB} and for GeMn containing ferromagnetic Ge$_3$Mn$_5$ nanocrystals\,\cite{Yu:2010_PRB}. When sweeping the magnetic field from -6\,T to 6\,T (stars -- upward arrow) and $vice\,versa$ (diamonds -- downward arrow), the maxima of the MR curves acquired at 2\,K are shifted by 120\,mT with respect to each other. This value corresponds to twice the coercive field at this temperature, as is indicated by the hysteresis shown in Fig.\,\ref{fig:MR2_1}(c), suggesting that until the magnetic moment of the NCs is aligned with the magnetic field, the electronic transport is sensitive to the spin polarization in the NCs.

The observed behavior of the MR as a function of temperature can be explained as follows: at temperatures above 50\,K, and due to phonon scattering, the spin life time of the spin-polarized conduction electrons in the $\gamma$'-Ga$_y$Fe$_{4-y}$N NCs is too short to span the distance between the NCs. However, for temperatures below 25\,K a spin-polarized current can be transported between the NCs, leading to a reduction of the MR. For temperatures around 10\,K, the $R_\mathrm{h}$ is too low for a hopping conduction between NCs to take place and hopping $via$ intermediate states can be ruled out, due to spin flip processes at the paramagnetic spins of the dilute Fe ions in the host matrix\,\cite{Zhao:1991_PRB}. At temperatures below 10\,K, $R_\mathrm{h}$ becomes sufficiently large for inter-NC-hopping to occur, increasing the hopping probability.

The MR of Sample B differs significantly from the one of the other investigated samples: it is positive down to 150\,K and it turns negative at 100\,K\,\cite{supplementary}. Due to the mentioned diffusion of dilute Fe from the layer containing the NCs into the GaN capping layer, the disorder of the system increases and the VRH conduction mechanism is likely to dominate already at temperatures as high as 100\,K. This conclusion is supported by the fact that a similar MR behavior is observed in semi-insulating dilute GaN:Fe layers, as reported in Fig.S3(d) of the Supplemental Material\,\cite{supplementary}. However, as the measured conductivity is a combination of the conductivities of the underlying GaN buffer, of the Ga$\delta$FeN layer containing the NCs and of the GaN(:Fe)capping layer, in this sample also the Arrhenius behavior from the buffer layer is significant in a wide temperature range.

The contribution of the embedded NCs to the MR in Sample A* is further tested by changing the direction of the applied magnetic field by the out-of-plane angle $\beta$ with respect to the surface normal from perpendicular ($\beta = 90^\circ$) to parallel ($\beta = 0^\circ$). The anisotropic magnetoresistance is defined as:

\begin{equation}
\mathrm{AMR} = \frac{\rho(\beta)-\rho(0^\circ)}{\rho(0^\circ)}.
\end{equation}

\begin{figure}
	\includegraphics[width=\columnwidth]{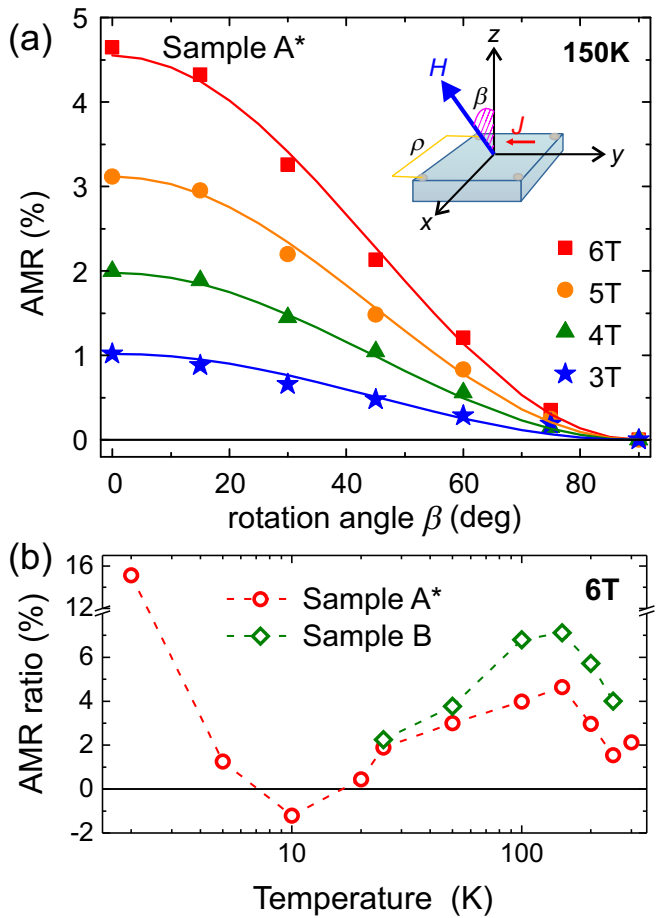}
	\caption{(a) AMR as a function of the out-of-plane angle $\beta$ between the magentic field and the surface normal (symbols) in Sample A* fitted by a $cos^2\beta$ function (solid lines). (c) AMR ratio as a function of temperature for Sample A* and Sample B.}
	\label{fig:AMR}
\end{figure}

All samples investigated and containing Ga$\delta$FeN show a positive AMR throughout the entire temperature range between 2\,K and 300\,K. The AMR is not observed in the GaN reference, pointing at the observed AMR as due to the $\gamma$'-Ga$_y$Fe$_{4-y}$N nanocrystals present in the other layers. The AMR acquired at 150\,K for Sample A* is reported in Figs.\,\ref{fig:AMR}(a). The AMR follows a $cos^2(\beta)$ behavior, consistent with the out-of-plane magnetic anisotropy of the ferromagnetic Ga$_y$Fe$_{4-y}$N NCs obtained by FMR. A positive AMR has been also reported for epitaxial thin films of $\gamma$'-Fe$_4$N grown on SrTiO$_3$(001) substrates\,\cite{Nikolaev:2003_APL}.

The evolution of the AMR ratio, defined as $(\rho_{\parallel}-\rho_{\perp})/\rho_{\perp}$, where $\rho_{\parallel}$ and $\rho_{\perp}$ are the resistivities for an applied field respectively parallel or perpendicular to the sample normal, as a function of temperature is presented in Fig.\,\ref{fig:AMR}(b) for all investigated samples. The highest AMR for Sample A* is 16\% at 2\,K, while it reaches values around (2-4)\% up to 300\,K. The AMR values for Sample B are slightly larger than those of Sample A*, pointing to a greater amount of ferromagnetic NCs in this sample. This is in accordance with the faster saturation observed in the magnetization of Sample B in Fig.\,\ref{fig:magnetic}(a). 

Anisotropic magnetoresistance is an effect generally dominated by spin-orbit interaction, which causes spin-mixing in the scattering processes of the conduction electrons with $3d$ orbitals, \textit{i.e.} \textit{s-d} scattering. When in $\gamma$'-Fe$_4$N the electron occupation changes as a function of the direction of the magnetic moment with respect to the cubic axes, the AMR is affected\,\cite{Tsunoda:2010_APEX}. In $\gamma$'-Fe$_4$N, the conduction electrons are preferentially scattered by the $3d$ orbitals when the magnetization is parallel to the easy axis, leading to an increased AMR when the current flows along this direction. According to the FMR measurements, the $\gamma$'-Ga$_y$Fe$_{4-y}$N nanocrystals embedded in GaN have a strong uniaxal magnetic anisotropy with three easy axes lying in the plane normal to the $c$-axis of GaN\,\cite{Grois:2014_Nanotech}. Therefore, when the magnetic field is applied parallel to the sample surface ($\beta = 0^\circ$), \textit{i.e.} parallel to the in-plane easy axis, the electron scattering is enhanced and hence, an increased resistance is observed. In contrast, when the field is applied normal to the surface ($\beta=90^\circ$), the electron scattering is reduced and the resistance diminishes.

 \section{Conclusions}

The magnetotransport in phase separated (Ga,Fe)N containing $\gamma$'-Ga$_y$Fe$_{4-y}$N NCs and grown by MOVPE on a GaN buffer deposited on $c$-sapphire is investigated.
The results show that, while the overall conduction mechanism in the studied layers occurs at the unintentionally \textit{n}-doped GaN buffer, the MR is significantly affected by the presence of the embedded $\gamma$'-Ga$_y$Fe$_{4-y}$N NCs. This is supported by the observed MR hysteresis at magnetic fields below 1\,T. The behavior of the MR in these layers can be described in terms of: (i) an Arrhenius-like mechanism for temperatures $\geq$50\,K, and (ii) an inter-nanocrystal hopping conduction, where the spin-polarized current is transported between NCs at temperatures below 25\,K in a regime in which phonon-scattering effects are not dominant. The hopping probability is increased at temperatures below 10\,K, where $R_\mathrm{h}$ becomes sufficiently large for inter-hopping to occur. 

In contrast to the negative AMR generally observed in $\gamma$-Fe$_4$N polycrystalline and epitaxial thin layers\,\cite{Ito:2012_JJAP,Tsunoda:2010_APEX}, the AMR of the $\gamma$'-Ga$_y$Fe$_{4-y}$N NCs embedded in GaN is positive with a high resistance state when a magnetic field is applied along the surface plane -- \textit{i.e.} parallel to the easy axis of the magnetization -- and with a low resistance state when the magnetic field is applied perpendicular to the surface. This is consistent with the in-plane shape anisotropy of the ferromagnetic NCs.

The observed values of the AMR ratio are $\sim$(2-3)\%  at RT, $i.e.$ significantly higher than the 0.17\% so far reported for $\gamma$'-Fe$_4$N thin films\,\cite{Nikolaev:2003_APL}, and open wide perspectives for the manipulation of AMR by external electric and magnetic fields -- previously reported for dilute (Ga,Mn)N \,\cite{Sztenkiel:2016_NComm} -- in these phase-separated systems.

\begin{acknowledgments}
This work was supported by the Austrian Science Fund (FWF) through the Elise-Richter Project No.V478, and projects P26830 and P24471, and the Austrian Exchange Service (\"{O}AD) project PL-01/2017, by the National Science Centre (Poland) [project OPUS (DEC-2013/09/B/ST3/04175)] and by the European Commission through the InTechFun (POIG.01.03.01-00-159/08) grant. The authors want to acknowledge C. Atteneder, P. Lindner and S. Wimmer for technical support and P. Dluzewski for his contribution to the HRTEM measurements. 
\end{acknowledgments}


\begin{thebibliography}{47}%
	\makeatletter
	\providecommand \@ifxundefined [1]{%
		\@ifx{#1\undefined}
	}%
	\providecommand \@ifnum [1]{%
		\ifnum #1\expandafter \@firstoftwo
		\else \expandafter \@secondoftwo
		\fi
	}%
	\providecommand \@ifx [1]{%
		\ifx #1\expandafter \@firstoftwo
		\else \expandafter \@secondoftwo
		\fi
	}%
	\providecommand \natexlab [1]{#1}%
	\providecommand \enquote  [1]{``#1''}%
	\providecommand \bibnamefont  [1]{#1}%
	\providecommand \bibfnamefont [1]{#1}%
	\providecommand \citenamefont [1]{#1}%
	\providecommand \href@noop [0]{\@secondoftwo}%
	\providecommand \href [0]{\begingroup \@sanitize@url \@href}%
	\providecommand \@href[1]{\@@startlink{#1}\@@href}%
	\providecommand \@@href[1]{\endgroup#1\@@endlink}%
	\providecommand \@sanitize@url [0]{\catcode `\\12\catcode `\$12\catcode
		`\&12\catcode `\#12\catcode `\^12\catcode `\_12\catcode `\%12\relax}%
	\providecommand \@@startlink[1]{}%
	\providecommand \@@endlink[0]{}%
	\providecommand \url  [0]{\begingroup\@sanitize@url \@url }%
	\providecommand \@url [1]{\endgroup\@href {#1}{\urlprefix }}%
	\providecommand \urlprefix  [0]{URL }%
	\providecommand \Eprint [0]{\href }%
	\providecommand \doibase [0]{http://dx.doi.org/}%
	\providecommand \selectlanguage [0]{\@gobble}%
	\providecommand \bibinfo  [0]{\@secondoftwo}%
	\providecommand \bibfield  [0]{\@secondoftwo}%
	\providecommand \translation [1]{[#1]}%
	\providecommand \BibitemOpen [0]{}%
	\providecommand \bibitemStop [0]{}%
	\providecommand \bibitemNoStop [0]{.\EOS\space}%
	\providecommand \EOS [0]{\spacefactor3000\relax}%
	\providecommand \BibitemShut  [1]{\csname bibitem#1\endcsname}%
	\let\auto@bib@innerbib\@empty
	\bibitem [{\citenamefont {Bonanni}\ and\ \citenamefont
		{Dietl}(2010)}]{Bonanni:2010_CSR}%
	\BibitemOpen
	\bibfield  {author} {\bibinfo {author} {\bibfnamefont {A.}~\bibnamefont
			{Bonanni}}\ and\ \bibinfo {author} {\bibfnamefont {T.}~\bibnamefont
			{Dietl}},\ }\href@noop {} {\bibfield  {journal} {\bibinfo  {journal} {Chem.
				Soc. Rev.}\ }\textbf {\bibinfo {volume} {39}},\ \bibinfo {pages} {528}
		(\bibinfo {year} {2010})}\BibitemShut {NoStop}%
	\bibitem [{\citenamefont {Sato}\ \emph {et~al.}(2010)\citenamefont {Sato},
		\citenamefont {Bergqvist}, \citenamefont {Kudrnovski}, \citenamefont
		{Dederichs}, \citenamefont {Eriksson}, \citenamefont {Turek}, \citenamefont
		{Sanyal}, \citenamefont {Bouzerar}, \citenamefont {Katayama-Yoshida},
		\citenamefont {Dinh}, \citenamefont {Fukushima}, \citenamefont {Kizaki},\
		and\ \citenamefont {Zeller}}]{Sato:2010_RMP}%
	\BibitemOpen
	\bibfield  {author} {\bibinfo {author} {\bibfnamefont {K.}~\bibnamefont
			{Sato}}, \bibinfo {author} {\bibfnamefont {L.}~\bibnamefont {Bergqvist}},
		\bibinfo {author} {\bibfnamefont {J.}~\bibnamefont {Kudrnovski}}, \bibinfo
		{author} {\bibfnamefont {P.~H.}\ \bibnamefont {Dederichs}}, \bibinfo {author}
		{\bibfnamefont {O.}~\bibnamefont {Eriksson}}, \bibinfo {author}
		{\bibfnamefont {I.}~\bibnamefont {Turek}}, \bibinfo {author} {\bibfnamefont
			{B.}~\bibnamefont {Sanyal}}, \bibinfo {author} {\bibfnamefont
			{G.}~\bibnamefont {Bouzerar}}, \bibinfo {author} {\bibfnamefont
			{H.}~\bibnamefont {Katayama-Yoshida}}, \bibinfo {author} {\bibfnamefont
			{V.~A.}\ \bibnamefont {Dinh}}, \bibinfo {author} {\bibfnamefont
			{T.}~\bibnamefont {Fukushima}}, \bibinfo {author} {\bibfnamefont
			{H.}~\bibnamefont {Kizaki}}, \ and\ \bibinfo {author} {\bibfnamefont
			{R.}~\bibnamefont {Zeller}},\ }\href@noop {} {\bibfield  {journal} {\bibinfo
			{journal} {Rev. Mod. Phys.}\ }\textbf {\bibinfo {volume} {82}},\ \bibinfo
		{pages} {1633} (\bibinfo {year} {2010})}\BibitemShut {NoStop}%
	\bibitem [{\citenamefont {Dietl}\ \emph {et~al.}(2015)\citenamefont {Dietl},
		\citenamefont {Sato}, \citenamefont {Fukushima}, \citenamefont {Bonanni},
		\citenamefont {Jamet}, \citenamefont {Barski}, \citenamefont {Kuroda},
		\citenamefont {Tanaka}, \citenamefont {Hai},\ and\ \citenamefont
		{Katayama-Yoshida}}]{Dietl:2015_RMP}%
	\BibitemOpen
	\bibfield  {author} {\bibinfo {author} {\bibfnamefont {T.}~\bibnamefont
			{Dietl}}, \bibinfo {author} {\bibfnamefont {K.}~\bibnamefont {Sato}},
		\bibinfo {author} {\bibfnamefont {T.}~\bibnamefont {Fukushima}}, \bibinfo
		{author} {\bibfnamefont {A.}~\bibnamefont {Bonanni}}, \bibinfo {author}
		{\bibfnamefont {M.}~\bibnamefont {Jamet}}, \bibinfo {author} {\bibfnamefont
			{A.}~\bibnamefont {Barski}}, \bibinfo {author} {\bibfnamefont
			{S.}~\bibnamefont {Kuroda}}, \bibinfo {author} {\bibfnamefont
			{M.}~\bibnamefont {Tanaka}}, \bibinfo {author} {\bibfnamefont {P.~N.}\
			\bibnamefont {Hai}}, \ and\ \bibinfo {author} {\bibfnamefont
			{H.}~\bibnamefont {Katayama-Yoshida}},\ }\href@noop {} {\bibfield  {journal}
		{\bibinfo  {journal} {Rev. Mod. Phys.}\ }\textbf {\bibinfo {volume} {87}},\
		\bibinfo {pages} {1311} (\bibinfo {year} {2015})}\BibitemShut {NoStop}%
	\bibitem [{\citenamefont {Coey}\ and\ \citenamefont
		{Smith}(1999)}]{Coey:1999_JMMM}%
	\BibitemOpen
	\bibfield  {author} {\bibinfo {author} {\bibfnamefont {J.}~\bibnamefont
			{Coey}}\ and\ \bibinfo {author} {\bibfnamefont {P.}~\bibnamefont {Smith}},\
	}\href@noop {} {\bibfield  {journal} {\bibinfo  {journal} {J. Magn. Magn.
				Mater.}\ }\textbf {\bibinfo {volume} {200}},\ \bibinfo {pages} {405}
		(\bibinfo {year} {1999})}\BibitemShut {NoStop}%
	\bibitem [{\citenamefont {Akiyama}\ \emph {et~al.}(2012)\citenamefont
		{Akiyama}, \citenamefont {Ohya}, \citenamefont {Hai},\ and\ \citenamefont
		{Tanaka}}]{Akiyama:2012_APL}%
	\BibitemOpen
	\bibfield  {author} {\bibinfo {author} {\bibfnamefont {R.}~\bibnamefont
			{Akiyama}}, \bibinfo {author} {\bibfnamefont {S.}~\bibnamefont {Ohya}},
		\bibinfo {author} {\bibfnamefont {P.~N.}\ \bibnamefont {Hai}}, \ and\
		\bibinfo {author} {\bibfnamefont {M.}~\bibnamefont {Tanaka}},\ }\href@noop {}
	{\bibfield  {journal} {\bibinfo  {journal} {J. Appl. Phys.}\ }\textbf
		{\bibinfo {volume} {111}},\ \bibinfo {pages} {063716} (\bibinfo {year}
		{2012})}\BibitemShut {NoStop}%
	\bibitem [{\citenamefont {Akinaga}\ \emph {et~al.}(1998)\citenamefont
		{Akinaga}, \citenamefont {Boeck}, \citenamefont {Borghs}, \citenamefont
		{Miyanishi}, \citenamefont {Asamitsu}, \citenamefont {Roy}, \citenamefont
		{Tomioka},\ and\ \citenamefont {Kuo}}]{Akinaga:1998_APL}%
	\BibitemOpen
	\bibfield  {author} {\bibinfo {author} {\bibfnamefont {H.}~\bibnamefont
			{Akinaga}}, \bibinfo {author} {\bibfnamefont {J.~D.}\ \bibnamefont {Boeck}},
		\bibinfo {author} {\bibfnamefont {G.}~\bibnamefont {Borghs}}, \bibinfo
		{author} {\bibfnamefont {S.}~\bibnamefont {Miyanishi}}, \bibinfo {author}
		{\bibfnamefont {A.}~\bibnamefont {Asamitsu}}, \bibinfo {author}
		{\bibfnamefont {W.~V.}\ \bibnamefont {Roy}}, \bibinfo {author} {\bibfnamefont
			{Y.}~\bibnamefont {Tomioka}}, \ and\ \bibinfo {author} {\bibfnamefont
			{L.}~\bibnamefont {Kuo}},\ }\href@noop {} {\bibfield  {journal} {\bibinfo
			{journal} {Appl. Phys. Lett.}\ }\textbf {\bibinfo {volume} {72}},\ \bibinfo
		{pages} {3368} (\bibinfo {year} {1998})}\BibitemShut {NoStop}%
	\bibitem [{\citenamefont {Michel}\ \emph {et~al.}(2008)\citenamefont {Michel},
		\citenamefont {Elm}, \citenamefont {Goldlucke}, \citenamefont {Baranovskii},
		\citenamefont {Thomas}, \citenamefont {Heimbrodt},\ and\ \citenamefont
		{Klar}}]{Michel:2008_APL}%
	\BibitemOpen
	\bibfield  {author} {\bibinfo {author} {\bibfnamefont {C.}~\bibnamefont
			{Michel}}, \bibinfo {author} {\bibfnamefont {M.~T.}\ \bibnamefont {Elm}},
		\bibinfo {author} {\bibfnamefont {B.}~\bibnamefont {Goldlucke}}, \bibinfo
		{author} {\bibfnamefont {S.~D.}\ \bibnamefont {Baranovskii}}, \bibinfo
		{author} {\bibfnamefont {P.}~\bibnamefont {Thomas}}, \bibinfo {author}
		{\bibfnamefont {W.}~\bibnamefont {Heimbrodt}}, \ and\ \bibinfo {author}
		{\bibfnamefont {P.~J.}\ \bibnamefont {Klar}},\ }\href@noop {} {\bibfield
		{journal} {\bibinfo  {journal} {Appl. Phys. Lett.}\ }\textbf {\bibinfo
			{volume} {92}},\ \bibinfo {pages} {223119} (\bibinfo {year}
		{2008})}\BibitemShut {NoStop}%
	\bibitem [{\citenamefont {Yu}\ \emph {et~al.}(2010)\citenamefont {Yu},
		\citenamefont {Jamet}, \citenamefont {Devillers}, \citenamefont {Barski},
		\citenamefont {Bayle-Guillemaud}, \citenamefont {Beign\'{e}}, \citenamefont
		{Rothman}, \citenamefont {Baltz},\ and\ \citenamefont
		{Cibert}}]{Yu:2010_PRB}%
	\BibitemOpen
	\bibfield  {author} {\bibinfo {author} {\bibfnamefont {I.-S.}\ \bibnamefont
			{Yu}}, \bibinfo {author} {\bibfnamefont {M.}~\bibnamefont {Jamet}}, \bibinfo
		{author} {\bibfnamefont {T.}~\bibnamefont {Devillers}}, \bibinfo {author}
		{\bibfnamefont {A.}~\bibnamefont {Barski}}, \bibinfo {author} {\bibfnamefont
			{P.}~\bibnamefont {Bayle-Guillemaud}}, \bibinfo {author} {\bibfnamefont
			{C.}~\bibnamefont {Beign\'{e}}}, \bibinfo {author} {\bibfnamefont
			{J.}~\bibnamefont {Rothman}}, \bibinfo {author} {\bibfnamefont
			{V.}~\bibnamefont {Baltz}}, \ and\ \bibinfo {author} {\bibfnamefont
			{J.}~\bibnamefont {Cibert}},\ }\href@noop {} {\bibfield  {journal} {\bibinfo
			{journal} {Phys. Rev. B.}\ }\textbf {\bibinfo {volume} {82}},\ \bibinfo
		{pages} {035308} (\bibinfo {year} {2010})}\BibitemShut {NoStop}%
	\bibitem [{\citenamefont {Hamieh}\ \emph {et~al.}(2015)\citenamefont {Hamieh},
		\citenamefont {Jedrecy}, \citenamefont {Hebert}, \citenamefont {Demaille},\
		and\ \citenamefont {Perriere}}]{Hamieh:2015_PRB}%
	\BibitemOpen
	\bibfield  {author} {\bibinfo {author} {\bibfnamefont {M.}~\bibnamefont
			{Hamieh}}, \bibinfo {author} {\bibfnamefont {N.}~\bibnamefont {Jedrecy}},
		\bibinfo {author} {\bibfnamefont {C.}~\bibnamefont {Hebert}}, \bibinfo
		{author} {\bibfnamefont {D.}~\bibnamefont {Demaille}}, \ and\ \bibinfo
		{author} {\bibfnamefont {J.}~\bibnamefont {Perriere}},\ }\href@noop {}
	{\bibfield  {journal} {\bibinfo  {journal} {Phys. Rev. B.}\ }\textbf
		{\bibinfo {volume} {92}},\ \bibinfo {pages} {155302} (\bibinfo {year}
		{2015})}\BibitemShut {NoStop}%
	\bibitem [{\citenamefont {Giraud}\ \emph {et~al.}(2004)\citenamefont {Giraud},
		\citenamefont {Kuroda}, \citenamefont {Marcet}, \citenamefont
		{Bellet-Amalric}, \citenamefont {Biquard}, \citenamefont {Barbara},
		\citenamefont {Fruchart}, \citenamefont {Ferrand}, \citenamefont {Cibert},\
		and\ \citenamefont {Mariette}}]{Giroud:2004_EPL}%
	\BibitemOpen
	\bibfield  {author} {\bibinfo {author} {\bibfnamefont {R.}~\bibnamefont
			{Giraud}}, \bibinfo {author} {\bibfnamefont {S.}~\bibnamefont {Kuroda}},
		\bibinfo {author} {\bibfnamefont {S.}~\bibnamefont {Marcet}}, \bibinfo
		{author} {\bibfnamefont {E.}~\bibnamefont {Bellet-Amalric}}, \bibinfo
		{author} {\bibfnamefont {X.}~\bibnamefont {Biquard}}, \bibinfo {author}
		{\bibfnamefont {B.}~\bibnamefont {Barbara}}, \bibinfo {author} {\bibfnamefont
			{D.}~\bibnamefont {Fruchart}}, \bibinfo {author} {\bibfnamefont
			{D.}~\bibnamefont {Ferrand}}, \bibinfo {author} {\bibfnamefont
			{J.}~\bibnamefont {Cibert}}, \ and\ \bibinfo {author} {\bibfnamefont
			{H.}~\bibnamefont {Mariette}},\ }\href@noop {} {\bibfield  {journal}
		{\bibinfo  {journal} {European Phys. Lett.}\ }\textbf {\bibinfo {volume}
			{65}},\ \bibinfo {pages} {553} (\bibinfo {year} {2004})}\BibitemShut
	{NoStop}%
	\bibitem [{\citenamefont {Kunert}\ \emph {et~al.}(2012)\citenamefont {Kunert},
		\citenamefont {Dobkowska}, \citenamefont {Li}, \citenamefont {Reuther},
		\citenamefont {Kruse}, \citenamefont {Figge}, \citenamefont {Jakiela},
		\citenamefont {Bonanni}, \citenamefont {Grenzer}, \citenamefont
		{Stefanowicz}, \citenamefont {von Borany}, \citenamefont {Sawicki},
		\citenamefont {Dietl},\ and\ \citenamefont {Hommel}}]{Kunert:2012_APL}%
	\BibitemOpen
	\bibfield  {author} {\bibinfo {author} {\bibfnamefont {G.}~\bibnamefont
			{Kunert}}, \bibinfo {author} {\bibfnamefont {S.}~\bibnamefont {Dobkowska}},
		\bibinfo {author} {\bibfnamefont {T.}~\bibnamefont {Li}}, \bibinfo {author}
		{\bibfnamefont {H.}~\bibnamefont {Reuther}}, \bibinfo {author} {\bibfnamefont
			{C.}~\bibnamefont {Kruse}}, \bibinfo {author} {\bibfnamefont
			{S.}~\bibnamefont {Figge}}, \bibinfo {author} {\bibfnamefont
			{R.}~\bibnamefont {Jakiela}}, \bibinfo {author} {\bibfnamefont
			{A.}~\bibnamefont {Bonanni}}, \bibinfo {author} {\bibfnamefont
			{J.}~\bibnamefont {Grenzer}}, \bibinfo {author} {\bibfnamefont
			{W.}~\bibnamefont {Stefanowicz}}, \bibinfo {author} {\bibfnamefont
			{J.}~\bibnamefont {von Borany}}, \bibinfo {author} {\bibfnamefont
			{M.}~\bibnamefont {Sawicki}}, \bibinfo {author} {\bibfnamefont
			{T.}~\bibnamefont {Dietl}}, \ and\ \bibinfo {author} {\bibfnamefont
			{D.}~\bibnamefont {Hommel}},\ }\href@noop {} {\bibfield  {journal} {\bibinfo
			{journal} {Appl. Phys. Lett.}\ }\textbf {\bibinfo {volume} {101}},\ \bibinfo
		{pages} {022413} (\bibinfo {year} {2012})}\BibitemShut {NoStop}%
	\bibitem [{\citenamefont {Bonanni}\ \emph {et~al.}(2008)\citenamefont
		{Bonanni}, \citenamefont {Navarro-Quezada}, \citenamefont {Li}, \citenamefont
		{Wegscheider}, \citenamefont {Mat\v{e}j}, \citenamefont {Hol\'{y}},
		\citenamefont {Lechner}, \citenamefont {Bauer}, \citenamefont {Rovezzi},
		\citenamefont {D'Acapito}, \citenamefont {Kiecana}, \citenamefont {Sawicki},\
		and\ \citenamefont {Dietl}}]{Bonanni:2008_PRL}%
	\BibitemOpen
	\bibfield  {author} {\bibinfo {author} {\bibfnamefont {A.}~\bibnamefont
			{Bonanni}}, \bibinfo {author} {\bibfnamefont {A.}~\bibnamefont
			{Navarro-Quezada}}, \bibinfo {author} {\bibfnamefont {T.}~\bibnamefont {Li}},
		\bibinfo {author} {\bibfnamefont {M.}~\bibnamefont {Wegscheider}}, \bibinfo
		{author} {\bibfnamefont {Z.}~\bibnamefont {Mat\v{e}j}}, \bibinfo {author}
		{\bibfnamefont {V.}~\bibnamefont {Hol\'{y}}}, \bibinfo {author}
		{\bibfnamefont {R.~T.}\ \bibnamefont {Lechner}}, \bibinfo {author}
		{\bibfnamefont {G.}~\bibnamefont {Bauer}}, \bibinfo {author} {\bibfnamefont
			{M.}~\bibnamefont {Rovezzi}}, \bibinfo {author} {\bibfnamefont
			{F.}~\bibnamefont {D'Acapito}}, \bibinfo {author} {\bibfnamefont
			{M.}~\bibnamefont {Kiecana}}, \bibinfo {author} {\bibfnamefont
			{M.}~\bibnamefont {Sawicki}}, \ and\ \bibinfo {author} {\bibfnamefont
			{T.}~\bibnamefont {Dietl}},\ }\href@noop {} {\bibfield  {journal} {\bibinfo
			{journal} {Phys. Rev. Lett.}\ }\textbf {\bibinfo {volume} {101}},\ \bibinfo
		{pages} {135502} (\bibinfo {year} {2008})}\BibitemShut {NoStop}%
	\bibitem [{\citenamefont {Bonanni}\ \emph {et~al.}(2007)\citenamefont
		{Bonanni}, \citenamefont {Kiecana}, \citenamefont {Simbrunner}, \citenamefont
		{Li}, \citenamefont {Sawicki}, \citenamefont {Wegscheider}, \citenamefont
		{Quast}, \citenamefont {Przybylinska}, \citenamefont {Navarro-Quezada},
		\citenamefont {Jakie\l{}a}, \citenamefont {Wolos}, \citenamefont {Jantsch},\
		and\ \citenamefont {Dietl}}]{Bonanni:2007_PRB}%
	\BibitemOpen
	\bibfield  {author} {\bibinfo {author} {\bibfnamefont {A.}~\bibnamefont
			{Bonanni}}, \bibinfo {author} {\bibfnamefont {M.}~\bibnamefont {Kiecana}},
		\bibinfo {author} {\bibfnamefont {C.}~\bibnamefont {Simbrunner}}, \bibinfo
		{author} {\bibfnamefont {T.}~\bibnamefont {Li}}, \bibinfo {author}
		{\bibfnamefont {M.}~\bibnamefont {Sawicki}}, \bibinfo {author} {\bibfnamefont
			{M.}~\bibnamefont {Wegscheider}}, \bibinfo {author} {\bibfnamefont
			{M.}~\bibnamefont {Quast}}, \bibinfo {author} {\bibfnamefont
			{H.}~\bibnamefont {Przybylinska}}, \bibinfo {author} {\bibfnamefont
			{A.}~\bibnamefont {Navarro-Quezada}}, \bibinfo {author} {\bibfnamefont
			{R.}~\bibnamefont {Jakie\l{}a}}, \bibinfo {author} {\bibfnamefont
			{A.}~\bibnamefont {Wolos}}, \bibinfo {author} {\bibfnamefont
			{W.}~\bibnamefont {Jantsch}}, \ and\ \bibinfo {author} {\bibfnamefont
			{T.}~\bibnamefont {Dietl}},\ }\href@noop {} {\bibfield  {journal} {\bibinfo
			{journal} {Phys. Rev. B}\ }\textbf {\bibinfo {volume} {75}},\ \bibinfo
		{pages} {125210} (\bibinfo {year} {2007})}\BibitemShut {NoStop}%
	\bibitem [{\citenamefont {Navarro-Quezada}\ \emph {et~al.}(2010)\citenamefont
		{Navarro-Quezada}, \citenamefont {Stefanowicz}, \citenamefont {Li},
		\citenamefont {Faina}, \citenamefont {Rovezzi}, \citenamefont {Lechner},
		\citenamefont {Devillers}, \citenamefont {d'Acapito}, \citenamefont {Bauer},
		\citenamefont {Sawicki}, \citenamefont {Dietl},\ and\ \citenamefont
		{Bonanni}}]{Navarro:2010_PRB}%
	\BibitemOpen
	\bibfield  {author} {\bibinfo {author} {\bibfnamefont {A.}~\bibnamefont
			{Navarro-Quezada}}, \bibinfo {author} {\bibfnamefont {W.}~\bibnamefont
			{Stefanowicz}}, \bibinfo {author} {\bibfnamefont {T.}~\bibnamefont {Li}},
		\bibinfo {author} {\bibfnamefont {B.}~\bibnamefont {Faina}}, \bibinfo
		{author} {\bibfnamefont {M.}~\bibnamefont {Rovezzi}}, \bibinfo {author}
		{\bibfnamefont {R.}~\bibnamefont {Lechner}}, \bibinfo {author} {\bibfnamefont
			{T.}~\bibnamefont {Devillers}}, \bibinfo {author} {\bibfnamefont
			{F.}~\bibnamefont {d'Acapito}}, \bibinfo {author} {\bibfnamefont
			{G.}~\bibnamefont {Bauer}}, \bibinfo {author} {\bibfnamefont
			{M.}~\bibnamefont {Sawicki}}, \bibinfo {author} {\bibfnamefont
			{T.}~\bibnamefont {Dietl}}, \ and\ \bibinfo {author} {\bibfnamefont
			{A.}~\bibnamefont {Bonanni}},\ }\href@noop {} {\bibfield  {journal} {\bibinfo
			{journal} {Phys. Rev. B}\ }\textbf {\bibinfo {volume} {81}},\ \bibinfo
		{pages} {205206} (\bibinfo {year} {2010})}\BibitemShut {NoStop}%
	\bibitem [{\citenamefont {Navarro-Quezada}\ \emph {et~al.}(2011)\citenamefont
		{Navarro-Quezada}, \citenamefont {Swacki}, \citenamefont {Stefanowicz},
		\citenamefont {Li}, \citenamefont {Grois}, \citenamefont {Devillers},
		\citenamefont {Rovezzi}, \citenamefont {Jakie\l{}a}, \citenamefont {Faina},
		\citenamefont {Majewski}, \citenamefont {Sawicki}, \citenamefont {Dietl},\
		and\ \citenamefont {Bonanni}}]{Navarro:2011_PRB}%
	\BibitemOpen
	\bibfield  {author} {\bibinfo {author} {\bibfnamefont {A.}~\bibnamefont
			{Navarro-Quezada}}, \bibinfo {author} {\bibfnamefont {N.~G.}\ \bibnamefont
			{Swacki}}, \bibinfo {author} {\bibfnamefont {W.}~\bibnamefont {Stefanowicz}},
		\bibinfo {author} {\bibfnamefont {T.}~\bibnamefont {Li}}, \bibinfo {author}
		{\bibfnamefont {A.}~\bibnamefont {Grois}}, \bibinfo {author} {\bibfnamefont
			{T.}~\bibnamefont {Devillers}}, \bibinfo {author} {\bibfnamefont
			{M.}~\bibnamefont {Rovezzi}}, \bibinfo {author} {\bibfnamefont
			{R.}~\bibnamefont {Jakie\l{}a}}, \bibinfo {author} {\bibfnamefont
			{B.}~\bibnamefont {Faina}}, \bibinfo {author} {\bibfnamefont
			{J.}~\bibnamefont {Majewski}}, \bibinfo {author} {\bibfnamefont
			{M.}~\bibnamefont {Sawicki}}, \bibinfo {author} {\bibfnamefont
			{T.}~\bibnamefont {Dietl}}, \ and\ \bibinfo {author} {\bibfnamefont
			{A.}~\bibnamefont {Bonanni}},\ }\href@noop {} {\bibfield  {journal} {\bibinfo
			{journal} {Phys. Rev. B}\ }\textbf {\bibinfo {volume} {84}},\ \bibinfo
		{pages} {155321} (\bibinfo {year} {2011})}\BibitemShut {NoStop}%
	\bibitem [{\citenamefont {Navarro-Quezada}\ \emph {et~al.}(2012)\citenamefont
		{Navarro-Quezada}, \citenamefont {Devillers}, \citenamefont {Li},\ and\
		\citenamefont {Bonanni}}]{Navarro:2012_APL}%
	\BibitemOpen
	\bibfield  {author} {\bibinfo {author} {\bibfnamefont {A.}~\bibnamefont
			{Navarro-Quezada}}, \bibinfo {author} {\bibfnamefont {T.}~\bibnamefont
			{Devillers}}, \bibinfo {author} {\bibfnamefont {T.}~\bibnamefont {Li}}, \
		and\ \bibinfo {author} {\bibfnamefont {A.}~\bibnamefont {Bonanni}},\
	}\href@noop {} {\bibfield  {journal} {\bibinfo  {journal} {Appl. Phys.
				Lett.}\ }\textbf {\bibinfo {volume} {101}},\ \bibinfo {pages} {081911}
		(\bibinfo {year} {2012})}\BibitemShut {NoStop}%
	\bibitem [{\citenamefont {Grois}\ \emph {et~al.}(2014)\citenamefont {Grois},
		\citenamefont {Devillers}, \citenamefont {Li},\ and\ \citenamefont
		{Bonanni}}]{Grois:2014_Nanotech}%
	\BibitemOpen
	\bibfield  {author} {\bibinfo {author} {\bibfnamefont {A.}~\bibnamefont
			{Grois}}, \bibinfo {author} {\bibfnamefont {T.}~\bibnamefont {Devillers}},
		\bibinfo {author} {\bibfnamefont {T.}~\bibnamefont {Li}}, \ and\ \bibinfo
		{author} {\bibfnamefont {A.}~\bibnamefont {Bonanni}},\ }\href@noop {}
	{\bibfield  {journal} {\bibinfo  {journal} {Nanotechnology}\ }\textbf
		{\bibinfo {volume} {25}},\ \bibinfo {pages} {1} (\bibinfo {year}
		{2014})}\BibitemShut {NoStop}%
	\bibitem [{\citenamefont {Houeben}\ \emph {et~al.}(2009)\citenamefont
		{Houeben}, \citenamefont {Burghaus},\ and\ \citenamefont
		{Dronskowski}}]{Houben:2009_ChemMat}%
	\BibitemOpen
	\bibfield  {author} {\bibinfo {author} {\bibfnamefont {A.}~\bibnamefont
			{Houeben}}, \bibinfo {author} {\bibfnamefont {J.}~\bibnamefont {Burghaus}}, \
		and\ \bibinfo {author} {\bibfnamefont {R.}~\bibnamefont {Dronskowski}},\
	}\href@noop {} {\bibfield  {journal} {\bibinfo  {journal} {Chem. Mater.}\
		}\textbf {\bibinfo {volume} {21}},\ \bibinfo {pages} {4332} (\bibinfo {year}
		{2009})}\BibitemShut {NoStop}%
	\bibitem [{\citenamefont {Burghaus}\ \emph {et~al.}(2011)\citenamefont
		{Burghaus}, \citenamefont {Sougrati}, \citenamefont {Moechel}, \citenamefont
		{Houben}, \citenamefont {Hermann},\ and\ \citenamefont
		{Dronskowski}}]{Burghaus:2011_JSSC}%
	\BibitemOpen
	\bibfield  {author} {\bibinfo {author} {\bibfnamefont {J.}~\bibnamefont
			{Burghaus}}, \bibinfo {author} {\bibfnamefont {M.}~\bibnamefont {Sougrati}},
		\bibinfo {author} {\bibfnamefont {A.}~\bibnamefont {Moechel}}, \bibinfo
		{author} {\bibfnamefont {A.}~\bibnamefont {Houben}}, \bibinfo {author}
		{\bibfnamefont {R.~P.}\ \bibnamefont {Hermann}}, \ and\ \bibinfo {author}
		{\bibfnamefont {R.}~\bibnamefont {Dronskowski}},\ }\href@noop {} {\bibfield
		{journal} {\bibinfo  {journal} {J. Solid State Chem.}\ }\textbf {\bibinfo
			{volume} {184}},\ \bibinfo {pages} {2315} (\bibinfo {year}
		{2011})}\BibitemShut {NoStop}%
	\bibitem [{\citenamefont {Jungwirth}\ \emph {et~al.}(2016)\citenamefont
		{Jungwirth}, \citenamefont {Mart\'{i}}, \citenamefont {Wadley},\ and\
		\citenamefont {Wunderlich}}]{Jungwirth:2016_NatNano}%
	\BibitemOpen
	\bibfield  {author} {\bibinfo {author} {\bibfnamefont {T.}~\bibnamefont
			{Jungwirth}}, \bibinfo {author} {\bibfnamefont {X.}~\bibnamefont
			{Mart\'{i}}}, \bibinfo {author} {\bibfnamefont {P.}~\bibnamefont {Wadley}}, \
		and\ \bibinfo {author} {\bibfnamefont {J.}~\bibnamefont {Wunderlich}},\
	}\href@noop {} {\bibfield  {journal} {\bibinfo  {journal} {Nat. Nanotech.}\
		}\textbf {\bibinfo {volume} {11}},\ \bibinfo {pages} {231} (\bibinfo {year}
		{2016})}\BibitemShut {NoStop}%
	\bibitem [{\citenamefont {Wadley}\ \emph {et~al.}(2016)\citenamefont {Wadley},
		\citenamefont {Howells}, \citenamefont {Zelezn\'{y}}, \citenamefont
		{Andrews}, \citenamefont {Hills}, \citenamefont {Campion}, \citenamefont
		{Nov\'{a}k}, \citenamefont {Olejn\'{i}k}, \citenamefont {Maccherozzi},
		\citenamefont {Dhesi}, \citenamefont {Martin}, \citenamefont {T.Wagner},
		\citenamefont {Wunderlich}, \citenamefont {Freimuth}, \citenamefont
		{Mokrousov}, \citenamefont {Kune\v{s}}, \citenamefont {Chauhan},
		\citenamefont {Grzybowski}, \citenamefont {Rushforth}, \citenamefont
		{Edmonds}, \citenamefont {Gallagher},\ and\ \citenamefont
		{Jungwirth}}]{Wadley:2016_Science}%
	\BibitemOpen
	\bibfield  {author} {\bibinfo {author} {\bibfnamefont {P.}~\bibnamefont
			{Wadley}}, \bibinfo {author} {\bibfnamefont {B.}~\bibnamefont {Howells}},
		\bibinfo {author} {\bibfnamefont {J.}~\bibnamefont {Zelezn\'{y}}}, \bibinfo
		{author} {\bibfnamefont {C.}~\bibnamefont {Andrews}}, \bibinfo {author}
		{\bibfnamefont {V.}~\bibnamefont {Hills}}, \bibinfo {author} {\bibfnamefont
			{R.}~\bibnamefont {Campion}}, \bibinfo {author} {\bibfnamefont
			{V.}~\bibnamefont {Nov\'{a}k}}, \bibinfo {author} {\bibfnamefont
			{K.}~\bibnamefont {Olejn\'{i}k}}, \bibinfo {author} {\bibfnamefont
			{F.}~\bibnamefont {Maccherozzi}}, \bibinfo {author} {\bibfnamefont
			{S.}~\bibnamefont {Dhesi}}, \bibinfo {author} {\bibfnamefont
			{S.}~\bibnamefont {Martin}}, \bibinfo {author} {\bibnamefont {T.Wagner}},
		\bibinfo {author} {\bibfnamefont {J.}~\bibnamefont {Wunderlich}}, \bibinfo
		{author} {\bibfnamefont {F.}~\bibnamefont {Freimuth}}, \bibinfo {author}
		{\bibfnamefont {Y.}~\bibnamefont {Mokrousov}}, \bibinfo {author}
		{\bibfnamefont {J.}~\bibnamefont {Kune\v{s}}}, \bibinfo {author}
		{\bibfnamefont {J.~S.}\ \bibnamefont {Chauhan}}, \bibinfo {author}
		{\bibfnamefont {M.}~\bibnamefont {Grzybowski}}, \bibinfo {author}
		{\bibfnamefont {A.~W.}\ \bibnamefont {Rushforth}}, \bibinfo {author}
		{\bibfnamefont {K.}~\bibnamefont {Edmonds}}, \bibinfo {author} {\bibfnamefont
			{B.~L.}\ \bibnamefont {Gallagher}}, \ and\ \bibinfo {author} {\bibfnamefont
			{T.}~\bibnamefont {Jungwirth}},\ }\href@noop {} {\bibfield  {journal}
		{\bibinfo  {journal} {Science}\ }\textbf {\bibinfo {volume} {351}},\ \bibinfo
		{pages} {587} (\bibinfo {year} {2016})}\BibitemShut {NoStop}%
	\bibitem [{\citenamefont {Sawicki}\ \emph {et~al.}(2011)\citenamefont
		{Sawicki}, \citenamefont {Stefanowicz},\ and\ \citenamefont
		{A.Ney}}]{Sawicki:2011_SST}%
	\BibitemOpen
	\bibfield  {author} {\bibinfo {author} {\bibfnamefont {M.}~\bibnamefont
			{Sawicki}}, \bibinfo {author} {\bibfnamefont {W.}~\bibnamefont
			{Stefanowicz}}, \ and\ \bibinfo {author} {\bibnamefont {A.Ney}},\ }\href@noop
	{} {\bibfield  {journal} {\bibinfo  {journal} {Semicond. Sci. Technol.}\
		}\textbf {\bibinfo {volume} {26}},\ \bibinfo {pages} {064006} (\bibinfo
		{year} {2011})}\BibitemShut {NoStop}%
	\bibitem [{\citenamefont {Gas}\ and\ \citenamefont
		{Sawicki}(2018)}]{Gas:2018_arxiv}%
	\BibitemOpen
	\bibfield  {author} {\bibinfo {author} {\bibfnamefont {K.}~\bibnamefont
			{Gas}}\ and\ \bibinfo {author} {\bibfnamefont {M.}~\bibnamefont {Sawicki}},\
	}\href@noop {} {\bibfield  {journal} {\bibinfo  {journal} {arXiv:1809.02346}\
		} (\bibinfo {year} {2018})}\BibitemShut {NoStop}%
	\bibitem [{sup()}]{supplementary}%
	\BibitemOpen
	\href@noop {} {}\bibinfo {note} {See Supplemental Material at ........for
		details on: the effect of annealing, x-ray photoemission spectroscopy
		analysis before and after annealing, electrical contacts, difference in
		magnetic moment between as-grown and annealed samples and MR of reference
		samples.}\BibitemShut {Stop}%
	\bibitem [{\citenamefont {Rovezzi}(2012)}]{Rovezzi:2012_arxiv}%
	\BibitemOpen
	\bibfield  {author} {\bibinfo {author} {\bibfnamefont {M.}~\bibnamefont
			{Rovezzi}},\ }\href@noop {} {\bibfield  {journal} {\bibinfo  {journal}
			{arXiv:1208.3420}\ } (\bibinfo {year} {2012})}\BibitemShut {NoStop}%
	\bibitem [{\citenamefont {Li}\ \emph {et~al.}(2008)\citenamefont {Li},
		\citenamefont {Simbrunner}, \citenamefont {Navarro-Quezada}, \citenamefont
		{Wegscheider}, \citenamefont {Quast}, \citenamefont {Litvinov}, \citenamefont
		{Gerthsen},\ and\ \citenamefont {Bonanni}}]{Li:2008_JCG}%
	\BibitemOpen
	\bibfield  {author} {\bibinfo {author} {\bibfnamefont {T.}~\bibnamefont
			{Li}}, \bibinfo {author} {\bibfnamefont {C.}~\bibnamefont {Simbrunner}},
		\bibinfo {author} {\bibfnamefont {A.}~\bibnamefont {Navarro-Quezada}},
		\bibinfo {author} {\bibfnamefont {M.}~\bibnamefont {Wegscheider}}, \bibinfo
		{author} {\bibfnamefont {M.}~\bibnamefont {Quast}}, \bibinfo {author}
		{\bibfnamefont {D.}~\bibnamefont {Litvinov}}, \bibinfo {author}
		{\bibfnamefont {D.}~\bibnamefont {Gerthsen}}, \ and\ \bibinfo {author}
		{\bibfnamefont {A.}~\bibnamefont {Bonanni}},\ }\href@noop {} {\bibfield
		{journal} {\bibinfo  {journal} {J. Cryst. Growth}\ }\textbf {\bibinfo
			{volume} {310}},\ \bibinfo {pages} {3294} (\bibinfo {year}
		{2008})}\BibitemShut {NoStop}%
	\bibitem [{\citenamefont {Patterson}(1939)}]{Patterson:1939_PR}%
	\BibitemOpen
	\bibfield  {author} {\bibinfo {author} {\bibfnamefont {A.}~\bibnamefont
			{Patterson}},\ }\href@noop {} {\bibfield  {journal} {\bibinfo  {journal}
			{Phys. Rev.}\ }\textbf {\bibinfo {volume} {56}},\ \bibinfo {pages} {978}
		(\bibinfo {year} {1939})}\BibitemShut {NoStop}%
	\bibitem [{\citenamefont {Pacuski}\ \emph {et~al.}(2008)\citenamefont
		{Pacuski}, \citenamefont {Kossacki}, \citenamefont {Ferrand}, \citenamefont
		{Golnik}, \citenamefont {Cibert}, \citenamefont {Wegscheider}, \citenamefont
		{Navarro-Quezada}, \citenamefont {Bonanni}, \citenamefont {Kiecana},
		\citenamefont {Sawicki},\ and\ \citenamefont {Dietl}}]{Pacuski:2008_PRL}%
	\BibitemOpen
	\bibfield  {author} {\bibinfo {author} {\bibfnamefont {W.}~\bibnamefont
			{Pacuski}}, \bibinfo {author} {\bibfnamefont {P.}~\bibnamefont {Kossacki}},
		\bibinfo {author} {\bibfnamefont {D.}~\bibnamefont {Ferrand}}, \bibinfo
		{author} {\bibfnamefont {A.}~\bibnamefont {Golnik}}, \bibinfo {author}
		{\bibfnamefont {J.}~\bibnamefont {Cibert}}, \bibinfo {author} {\bibfnamefont
			{M.}~\bibnamefont {Wegscheider}}, \bibinfo {author} {\bibfnamefont
			{A.}~\bibnamefont {Navarro-Quezada}}, \bibinfo {author} {\bibfnamefont
			{A.}~\bibnamefont {Bonanni}}, \bibinfo {author} {\bibfnamefont
			{M.}~\bibnamefont {Kiecana}}, \bibinfo {author} {\bibfnamefont
			{M.}~\bibnamefont {Sawicki}}, \ and\ \bibinfo {author} {\bibfnamefont
			{T.}~\bibnamefont {Dietl}},\ }\href@noop {} {\bibfield  {journal} {\bibinfo
			{journal} {Phys. Rev. Lett.}\ }\textbf {\bibinfo {volume} {100}},\ \bibinfo
		{pages} {037204} (\bibinfo {year} {2008})}\BibitemShut {NoStop}%
	\bibitem [{\citenamefont {Nie}\ \emph {et~al.}(2003)\citenamefont {Nie},
		\citenamefont {Xu}, \citenamefont {Ong}, \citenamefont {Zhan}, \citenamefont
		{D.X.Li},\ and\ \citenamefont {Wang}}]{Nie:2003_TSF}%
	\BibitemOpen
	\bibfield  {author} {\bibinfo {author} {\bibfnamefont {H.}~\bibnamefont
			{Nie}}, \bibinfo {author} {\bibfnamefont {S.}~\bibnamefont {Xu}}, \bibinfo
		{author} {\bibfnamefont {C.}~\bibnamefont {Ong}}, \bibinfo {author}
		{\bibfnamefont {Q.}~\bibnamefont {Zhan}}, \bibinfo {author} {\bibnamefont
			{D.X.Li}}, \ and\ \bibinfo {author} {\bibfnamefont {J.}~\bibnamefont
			{Wang}},\ }\href@noop {} {\bibfield  {journal} {\bibinfo  {journal} {Thin
				Solid Films}\ }\textbf {\bibinfo {volume} {440}},\ \bibinfo {pages} {35}
		(\bibinfo {year} {2003})}\BibitemShut {NoStop}%
	\bibitem [{\citenamefont {Wu}\ \emph {et~al.}(2004)\citenamefont {Wu},
		\citenamefont {Zhong}, \citenamefont {Jiang}, \citenamefont {Tang},
		\citenamefont {Zou},\ and\ \citenamefont {Du}}]{Wu:2004_JMMM}%
	\BibitemOpen
	\bibfield  {author} {\bibinfo {author} {\bibfnamefont {X.}~\bibnamefont
			{Wu}}, \bibinfo {author} {\bibfnamefont {W.}~\bibnamefont {Zhong}}, \bibinfo
		{author} {\bibfnamefont {H.}~\bibnamefont {Jiang}}, \bibinfo {author}
		{\bibfnamefont {N.}~\bibnamefont {Tang}}, \bibinfo {author} {\bibfnamefont
			{W.}~\bibnamefont {Zou}}, \ and\ \bibinfo {author} {\bibfnamefont
			{Y.}~\bibnamefont {Du}},\ }\href@noop {} {\bibfield  {journal} {\bibinfo
			{journal} {J. Magn. Magn. Mater.}\ }\textbf {\bibinfo {volume} {281}},\
		\bibinfo {pages} {77} (\bibinfo {year} {2004})}\BibitemShut {NoStop}%
	\bibitem [{\citenamefont {Zhang}\ \emph {et~al.}(1998)\citenamefont {Zhang},
		\citenamefont {Klabunde}, \citenamefont {Sorensen},\ and\ \citenamefont
		{Hadjipanayis}}]{Zhang:1998_PRB}%
	\BibitemOpen
	\bibfield  {author} {\bibinfo {author} {\bibfnamefont {D.}~\bibnamefont
			{Zhang}}, \bibinfo {author} {\bibfnamefont {K.~J.}\ \bibnamefont {Klabunde}},
		\bibinfo {author} {\bibfnamefont {C.~M.}\ \bibnamefont {Sorensen}}, \ and\
		\bibinfo {author} {\bibfnamefont {G.~C.}\ \bibnamefont {Hadjipanayis}},\
	}\href@noop {} {\bibfield  {journal} {\bibinfo  {journal} {Phys. Rev. B}\
		}\textbf {\bibinfo {volume} {58}},\ \bibinfo {pages} {14167} (\bibinfo {year}
		{1998})}\BibitemShut {NoStop}%
	\bibitem [{\citenamefont {Heikman}\ \emph {et~al.}(2002)\citenamefont
		{Heikman}, \citenamefont {Keller}, \citenamefont {DenBaars},\ and\
		\citenamefont {Mishra}}]{Heikman:2002_APL}%
	\BibitemOpen
	\bibfield  {author} {\bibinfo {author} {\bibfnamefont {S.}~\bibnamefont
			{Heikman}}, \bibinfo {author} {\bibfnamefont {S.}~\bibnamefont {Keller}},
		\bibinfo {author} {\bibfnamefont {S.~P.}\ \bibnamefont {DenBaars}}, \ and\
		\bibinfo {author} {\bibfnamefont {U.~K.}\ \bibnamefont {Mishra}},\
	}\href@noop {} {\bibfield  {journal} {\bibinfo  {journal} {Appl. Phys.
				Lett.}\ }\textbf {\bibinfo {volume} {81}},\ \bibinfo {pages} {439} (\bibinfo
		{year} {2002})}\BibitemShut {NoStop}%
	\bibitem [{\citenamefont {Goetz}\ \emph {et~al.}(1996)\citenamefont {Goetz},
		\citenamefont {Johnson}, \citenamefont {Chen}, \citenamefont {Liu},
		\citenamefont {Kuo},\ and\ \citenamefont {Imler}}]{Goetz:1996_APL}%
	\BibitemOpen
	\bibfield  {author} {\bibinfo {author} {\bibfnamefont {W.}~\bibnamefont
			{Goetz}}, \bibinfo {author} {\bibfnamefont {N.}~\bibnamefont {Johnson}},
		\bibinfo {author} {\bibfnamefont {C.}~\bibnamefont {Chen}}, \bibinfo {author}
		{\bibfnamefont {H.}~\bibnamefont {Liu}}, \bibinfo {author} {\bibfnamefont
			{C.}~\bibnamefont {Kuo}}, \ and\ \bibinfo {author} {\bibfnamefont
			{W.}~\bibnamefont {Imler}},\ }\href@noop {} {\bibfield  {journal} {\bibinfo
			{journal} {Appl. Phys. Lett.}\ }\textbf {\bibinfo {volume} {68}},\ \bibinfo
		{pages} {3144} (\bibinfo {year} {1996})}\BibitemShut {NoStop}%
	\bibitem [{\citenamefont {Yildiz}\ \emph {et~al.}(2010)\citenamefont {Yildiz},
		\citenamefont {Lisesivdin}, \citenamefont {Kasap}, \citenamefont {Ozcelik},
		\citenamefont {Ozbay},\ and\ \citenamefont {Balkan}}]{Yildiz:2010_APA}%
	\BibitemOpen
	\bibfield  {author} {\bibinfo {author} {\bibfnamefont {A.}~\bibnamefont
			{Yildiz}}, \bibinfo {author} {\bibfnamefont {S.}~\bibnamefont {Lisesivdin}},
		\bibinfo {author} {\bibfnamefont {M.}~\bibnamefont {Kasap}}, \bibinfo
		{author} {\bibfnamefont {S.}~\bibnamefont {Ozcelik}}, \bibinfo {author}
		{\bibfnamefont {E.}~\bibnamefont {Ozbay}}, \ and\ \bibinfo {author}
		{\bibfnamefont {N.}~\bibnamefont {Balkan}},\ }\href@noop {} {\bibfield
		{journal} {\bibinfo  {journal} {Appl. Phys. A}\ }\textbf {\bibinfo {volume}
			{98}},\ \bibinfo {pages} {557} (\bibinfo {year} {2010})}\BibitemShut
	{NoStop}%
	\bibitem [{\citenamefont {Pollak}(2002)}]{Pollak:2002_pssb}%
	\BibitemOpen
	\bibfield  {author} {\bibinfo {author} {\bibfnamefont {M.}~\bibnamefont
			{Pollak}},\ }\href@noop {} {\bibfield  {journal} {\bibinfo  {journal} {Phys.
				Status Solidi B}\ }\textbf {\bibinfo {volume} {230}},\ \bibinfo {pages} {295}
		(\bibinfo {year} {2002})}\BibitemShut {NoStop}%
	\bibitem [{\citenamefont {Mott}(1969)}]{Mott:1969}%
	\BibitemOpen
	\bibfield  {author} {\bibinfo {author} {\bibfnamefont {N.}~\bibnamefont
			{Mott}},\ }\href@noop {} {\bibfield  {journal} {\bibinfo  {journal} {Philos.
				Mag.}\ }\textbf {\bibinfo {volume} {19}},\ \bibinfo {pages} {835} (\bibinfo
		{year} {1969})}\BibitemShut {NoStop}%
	\bibitem [{\citenamefont {Bedoya-Pinto}\ \emph {et~al.}(2009)\citenamefont
		{Bedoya-Pinto}, \citenamefont {Malindretos}, \citenamefont {Roever},
		\citenamefont {Mai},\ and\ \citenamefont {Rizzi}}]{Pinto:2009_PRB}%
	\BibitemOpen
	\bibfield  {author} {\bibinfo {author} {\bibfnamefont {A.}~\bibnamefont
			{Bedoya-Pinto}}, \bibinfo {author} {\bibfnamefont {J.}~\bibnamefont
			{Malindretos}}, \bibinfo {author} {\bibfnamefont {M.}~\bibnamefont {Roever}},
		\bibinfo {author} {\bibfnamefont {D.~D.}\ \bibnamefont {Mai}}, \ and\
		\bibinfo {author} {\bibfnamefont {A.}~\bibnamefont {Rizzi}},\ }\href@noop {}
	{\bibfield  {journal} {\bibinfo  {journal} {Phys. Rev. B.}\ }\textbf
		{\bibinfo {volume} {80}},\ \bibinfo {pages} {195208} (\bibinfo {year}
		{2009})}\BibitemShut {NoStop}%
	\bibitem [{\citenamefont {Zaremba}(1992)}]{Zaremba:1992_PRB}%
	\BibitemOpen
	\bibfield  {author} {\bibinfo {author} {\bibfnamefont {E.}~\bibnamefont
			{Zaremba}},\ }\href@noop {} {\bibfield  {journal} {\bibinfo  {journal} {Phys.
				Rev. B}\ }\textbf {\bibinfo {volume} {45}},\ \bibinfo {pages} {14143}
		(\bibinfo {year} {1992})}\BibitemShut {NoStop}%
	\bibitem [{\citenamefont {Zhao}\ \emph {et~al.}(1991)\citenamefont {Zhao},
		\citenamefont {Spivak}, \citenamefont {Gelfand},\ and\ \citenamefont
		{Feng}}]{Zhao:1991_PRB}%
	\BibitemOpen
	\bibfield  {author} {\bibinfo {author} {\bibfnamefont {H.}~\bibnamefont
			{Zhao}}, \bibinfo {author} {\bibfnamefont {B.}~\bibnamefont {Spivak}},
		\bibinfo {author} {\bibfnamefont {M.}~\bibnamefont {Gelfand}}, \ and\
		\bibinfo {author} {\bibfnamefont {S.}~\bibnamefont {Feng}},\ }\href@noop {}
	{\bibfield  {journal} {\bibinfo  {journal} {Phys. Rev. B}\ }\textbf {\bibinfo
			{volume} {44}},\ \bibinfo {pages} {10760} (\bibinfo {year}
		{1991})}\BibitemShut {NoStop}%
	\bibitem [{\citenamefont {Gantmakher}\ \emph {et~al.}(1996)\citenamefont
		{Gantmakher}, \citenamefont {Golubkov}, \citenamefont {Lok},\ and\
		\citenamefont {Geim}}]{Gantmakher:1996_JETP}%
	\BibitemOpen
	\bibfield  {author} {\bibinfo {author} {\bibfnamefont {V.}~\bibnamefont
			{Gantmakher}}, \bibinfo {author} {\bibfnamefont {M.}~\bibnamefont
			{Golubkov}}, \bibinfo {author} {\bibfnamefont {J.}~\bibnamefont {Lok}}, \
		and\ \bibinfo {author} {\bibfnamefont {A.}~\bibnamefont {Geim}},\ }\href@noop
	{} {\bibfield  {journal} {\bibinfo  {journal} {J. Exp.Theor. Phys.}\ }\textbf
		{\bibinfo {volume} {82}},\ \bibinfo {pages} {951} (\bibinfo {year}
		{1996})}\BibitemShut {NoStop}%
	\bibitem [{\citenamefont {Etin-Wohlman}\ \emph {et~al.}(1989)\citenamefont
		{Etin-Wohlman}, \citenamefont {Imry},\ and\ \citenamefont
		{Sivan}}]{Entin:1989_PRB}%
	\BibitemOpen
	\bibfield  {author} {\bibinfo {author} {\bibfnamefont {O.}~\bibnamefont
			{Etin-Wohlman}}, \bibinfo {author} {\bibfnamefont {Y.}~\bibnamefont {Imry}},
		\ and\ \bibinfo {author} {\bibfnamefont {U.}~\bibnamefont {Sivan}},\
	}\href@noop {} {\bibfield  {journal} {\bibinfo  {journal} {Phys.Rev. B}\
		}\textbf {\bibinfo {volume} {40}},\ \bibinfo {pages} {8342} (\bibinfo {year}
		{1989})}\BibitemShut {NoStop}%
	\bibitem [{\citenamefont {Abdia}\ \emph {et~al.}(2009)\citenamefont {Abdia},
		\citenamefont {Kaaouachi}, \citenamefont {Nafidi}, \citenamefont
		{Biskupski},\ and\ \citenamefont {Hemine}}]{Abdia:2009_SSE}%
	\BibitemOpen
	\bibfield  {author} {\bibinfo {author} {\bibfnamefont {R.}~\bibnamefont
			{Abdia}}, \bibinfo {author} {\bibfnamefont {A.}~\bibnamefont {Kaaouachi}},
		\bibinfo {author} {\bibfnamefont {A.}~\bibnamefont {Nafidi}}, \bibinfo
		{author} {\bibfnamefont {G.}~\bibnamefont {Biskupski}}, \ and\ \bibinfo
		{author} {\bibfnamefont {J.}~\bibnamefont {Hemine}},\ }\href@noop {}
	{\bibfield  {journal} {\bibinfo  {journal} {Solid State Electron.}\ }\textbf
		{\bibinfo {volume} {53}},\ \bibinfo {pages} {469} (\bibinfo {year}
		{2009})}\BibitemShut {NoStop}%
	\bibitem [{\citenamefont {Zhang}\ \emph {et~al.}(1992)\citenamefont {Zhang},
		\citenamefont {Dai},\ and\ \citenamefont {Sarachik}}]{Zhang:1992_PRB}%
	\BibitemOpen
	\bibfield  {author} {\bibinfo {author} {\bibfnamefont {Y.}~\bibnamefont
			{Zhang}}, \bibinfo {author} {\bibfnamefont {P.}~\bibnamefont {Dai}}, \ and\
		\bibinfo {author} {\bibfnamefont {M.}~\bibnamefont {Sarachik}},\ }\href@noop
	{} {\bibfield  {journal} {\bibinfo  {journal} {Phys. Rev. B}\ }\textbf
		{\bibinfo {volume} {45}},\ \bibinfo {pages} {9473} (\bibinfo {year}
		{1992})}\BibitemShut {NoStop}%
	\bibitem [{\citenamefont {Nikolaev}\ \emph {et~al.}(2003)\citenamefont
		{Nikolaev}, \citenamefont {Krivorotov}, \citenamefont {Dahlberg},
		\citenamefont {Vas'ko}, \citenamefont {Urazdhin}, \citenamefont {Loloee},\
		and\ \citenamefont {Pratt}}]{Nikolaev:2003_APL}%
	\BibitemOpen
	\bibfield  {author} {\bibinfo {author} {\bibfnamefont {K.}~\bibnamefont
			{Nikolaev}}, \bibinfo {author} {\bibfnamefont {I.}~\bibnamefont
			{Krivorotov}}, \bibinfo {author} {\bibfnamefont {E.}~\bibnamefont
			{Dahlberg}}, \bibinfo {author} {\bibfnamefont {V.}~\bibnamefont {Vas'ko}},
		\bibinfo {author} {\bibfnamefont {S.}~\bibnamefont {Urazdhin}}, \bibinfo
		{author} {\bibfnamefont {R.}~\bibnamefont {Loloee}}, \ and\ \bibinfo {author}
		{\bibfnamefont {W.}~\bibnamefont {Pratt}},\ }\href@noop {} {\bibfield
		{journal} {\bibinfo  {journal} {Appl. Phys. Lett.}\ }\textbf {\bibinfo
			{volume} {82}},\ \bibinfo {pages} {98} (\bibinfo {year} {2003})}\BibitemShut
	{NoStop}%
	\bibitem [{\citenamefont {Tsunoda}\ \emph {et~al.}(2010)\citenamefont
		{Tsunoda}, \citenamefont {Takahashi}, \citenamefont {Kokado}, \citenamefont
		{Komasaki}, \citenamefont {Sakuma},\ and\ \citenamefont
		{Takahashi}}]{Tsunoda:2010_APEX}%
	\BibitemOpen
	\bibfield  {author} {\bibinfo {author} {\bibfnamefont {M.}~\bibnamefont
			{Tsunoda}}, \bibinfo {author} {\bibfnamefont {H.}~\bibnamefont {Takahashi}},
		\bibinfo {author} {\bibfnamefont {S.}~\bibnamefont {Kokado}}, \bibinfo
		{author} {\bibfnamefont {Y.}~\bibnamefont {Komasaki}}, \bibinfo {author}
		{\bibfnamefont {A.}~\bibnamefont {Sakuma}}, \ and\ \bibinfo {author}
		{\bibfnamefont {M.}~\bibnamefont {Takahashi}},\ }\href@noop {} {\bibfield
		{journal} {\bibinfo  {journal} {Appl. Phys. Express}\ }\textbf {\bibinfo
			{volume} {3}},\ \bibinfo {pages} {113003} (\bibinfo {year}
		{2010})}\BibitemShut {NoStop}%
	\bibitem [{\citenamefont {Ito}\ \emph {et~al.}(2004)\citenamefont {Ito},
		\citenamefont {Kabara}, \citenamefont {Takahashi}, \citenamefont {Sanai},
		\citenamefont {Toko}, \citenamefont {Suemasu},\ and\ \citenamefont
		{Tsunoda}}]{Ito:2012_JJAP}%
	\BibitemOpen
	\bibfield  {author} {\bibinfo {author} {\bibfnamefont {K.}~\bibnamefont
			{Ito}}, \bibinfo {author} {\bibfnamefont {K.}~\bibnamefont {Kabara}},
		\bibinfo {author} {\bibfnamefont {H.}~\bibnamefont {Takahashi}}, \bibinfo
		{author} {\bibfnamefont {T.}~\bibnamefont {Sanai}}, \bibinfo {author}
		{\bibfnamefont {K.}~\bibnamefont {Toko}}, \bibinfo {author} {\bibfnamefont
			{T.}~\bibnamefont {Suemasu}}, \ and\ \bibinfo {author} {\bibfnamefont
			{M.}~\bibnamefont {Tsunoda}},\ }\href@noop {} {\bibfield  {journal} {\bibinfo
			{journal} {Jap. J. Appl. Phys.}\ }\textbf {\bibinfo {volume} {51}},\
		\bibinfo {pages} {068001} (\bibinfo {year} {2004})}\BibitemShut {NoStop}%
	\bibitem [{\citenamefont {Sztenkiel}\ \emph {et~al.}(2016)\citenamefont
		{Sztenkiel}, \citenamefont {Foltyn}, \citenamefont {Mazur}, \citenamefont
		{Adhikari}, \citenamefont {Kosiel}, \citenamefont {Gas}, \citenamefont
		{Zgirski}, \citenamefont {Kruszka}, \citenamefont {Jakie\l{}a}, \citenamefont
		{Li}, \citenamefont {Piotrowska}, \citenamefont {Bonanni}, \citenamefont
		{Sawicki},\ and\ \citenamefont {Dietl}}]{Sztenkiel:2016_NComm}%
	\BibitemOpen
	\bibfield  {author} {\bibinfo {author} {\bibfnamefont {D.}~\bibnamefont
			{Sztenkiel}}, \bibinfo {author} {\bibfnamefont {M.}~\bibnamefont {Foltyn}},
		\bibinfo {author} {\bibfnamefont {G.}~\bibnamefont {Mazur}}, \bibinfo
		{author} {\bibfnamefont {R.}~\bibnamefont {Adhikari}}, \bibinfo {author}
		{\bibfnamefont {K.}~\bibnamefont {Kosiel}}, \bibinfo {author} {\bibfnamefont
			{K.}~\bibnamefont {Gas}}, \bibinfo {author} {\bibfnamefont {M.}~\bibnamefont
			{Zgirski}}, \bibinfo {author} {\bibfnamefont {R.}~\bibnamefont {Kruszka}},
		\bibinfo {author} {\bibfnamefont {R.}~\bibnamefont {Jakie\l{}a}}, \bibinfo
		{author} {\bibfnamefont {T.}~\bibnamefont {Li}}, \bibinfo {author}
		{\bibfnamefont {A.}~\bibnamefont {Piotrowska}}, \bibinfo {author}
		{\bibfnamefont {A.}~\bibnamefont {Bonanni}}, \bibinfo {author} {\bibfnamefont
			{M.}~\bibnamefont {Sawicki}}, \ and\ \bibinfo {author} {\bibfnamefont
			{T.}~\bibnamefont {Dietl}},\ }\href@noop {} {\bibfield  {journal} {\bibinfo
			{journal} {Nat. Commun.}\ }\textbf {\bibinfo {volume} {7}},\ \bibinfo {pages}
		{13232} (\bibinfo {year} {2016})}\BibitemShut {NoStop}%
\end{thebibliography}
%

\end{document}